\documentclass[fleqn,usenatbib]{mnras}

\usepackage{newtxtext,newtxmath}

\usepackage[T1]{fontenc}
\usepackage{ae,aecompl}
\usepackage{soul}

\usepackage[usenames]{xcolor}
\usepackage[utf8]{inputenc}

\usepackage{amsmath}
\usepackage{graphicx}	
\usepackage{amssymb}	
\usepackage{natbib}
\usepackage{hyperref}
\usepackage{xfrac}

\DeclareMathOperator{\subthermal}{sub-thermal}
\DeclareMathOperator{\superthermal}{super-thermal}
\DeclareMathOperator{\repulsionlimited}{repulsion-limited}
\DeclareMathOperator{\consumptionlimited}{consumption-limited}
\DeclareMathAlphabet{\mathpzc}{OT1}{pzc}{m}{it}\definecolor{purple}{RGB}{160,32,240}

\global\long\def\pd#1#2{\frac{\partial#1}{\partial#2}}%

\title[Consumption and Repulsion]{How Consumption and Repulsion Set
  Planetary Gap Depths and the Final Masses of Gas Giants}

\author[Rosenthal et al.]{M.~M.~Rosenthal$^{1\href{https://orcid.org/0000-0003-3938-3099}{\includegraphics[scale=0.4]{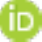}}}$\thanks{E-mail: mmrosent@ucsc.edu}, 
E.~I.~Chiang$^{2,3\href{https://orcid.org/0000-0002-6246-2310}{\includegraphics[scale=0.4]{orcid.pdf}}}$,
S.~Ginzburg$^{2}$,
R.~A.~Murray-Clay$^{1\href{https://orcid.org/0000-0001-5061-0462}{\includegraphics[scale=0.4]{orcid.pdf}}}$
\\
$^{1}$ Department of Astronomy and Astrophysics, University of California, Santa Cruz, CA 95064, USA\\
$^{2}$ Department of Astronomy, University of California at Berkeley, CA 94720, USA\\
$^{3}$ Department of Earth and Planetary Science, University of California, Berkeley, CA 94720, USA
}

\date{Released \today}

\pubyear{2020}

\begin{document}
\label{firstpage}
\pagerange{\pageref{firstpage}--\pageref{lastpage}}
\maketitle

\begin{abstract}
Planets open gaps in discs.
Gap opening is typically modeled by considering the planetary Lindblad torque which repels disc gas away from the planet's orbit. But gaps also clear because the planet consumes local material. We 
present a simple, easy-to-use, analytic framework for
calculating 
how gaps deplete and how the disc's structure as a whole changes by the combined action of Lindblad repulsion and planetary consumption. 
The final mass to which a gap-embedded gas giant grows is derived in tandem. The analytics are tested against 1D
numerical experiments and calibrated using published multi-dimensional
simulations.
In viscous alpha discs, the planet, while clearing a gap,  initially accretes practically all of the 
gas that tries to diffuse past, rapidly achieving super-Jupiter if not brown dwarf status. By contrast, in inviscid discs---that may still accrete onto their central stars by, say, magnetized winds---planets open deep, repulsion-dominated gaps. Then only a small fraction of the disc accretion flow is diverted onto the planet, which grows to a fraction of a Jupiter mass. Transitional disc cavities might be cleared by families of such low-mass objects opening inviscid,  repulsion-dominated, overlapping gaps which allow most
of the outer disc gas to flow unimpeded onto host stars.
\end{abstract}


\section{Introduction}\label{sec:intro}
Annular gaps in protoplanetary discs
are often attributed to embedded planets. 
The interpretation stems from the theory
of satellite-disc interactions that successfully
predicted the existence of shepherd moons in planetary
rings (e.g., \citealt{gt_1982}). 
Satellites in rings, and by analogy planets in discs,
repel material away from their orbits
as the waves they excite at Lindblad resonances 
dissipate and impart angular momentum to the ambient medium (see also
\citealt{gr_2001}; \citealt{gs_2018}).
The repulsive, gap-forming planetary Lindblad torque
competes against the disc's viscous torque 
which diffuses material back into the gap.

Most studies of protoplanetary disc gaps
concentrate exclusively on the Lindblad and viscous 
torques (e.g., \citealt{fsc_2014}; \citealt{kmt_2015}; \citealt{zzh_2018}) and neglect how gaps can also deepen 
because embedded planets consume local disc gas. 
Exceptions include, e.g., \citet{znh_2011}, \citet{dk_2015,dk_2017}, and \citet{mfv_2019}, 
whose numerical simulations of planet-disc interactions
allow for both planetary accretion and planetary torques. 
Our aim here is to give an elementary and analytic accounting of both effects:
to understand, for planets on fixed circular orbits,
how Lindblad repulsion and planetary consumption
combine to set gap depths.
This is a two-way feedback problem---planetary accretion
affects the gas density inside the gap, but the density
inside the gap determines the rate of planetary accretion 
(\citealt{gc_2019a}, \citeyear{gc_2019b}). 
Accordingly we will calculate how gas giants grow
in tandem with their deepening gaps.
Much of our analytic framework is the same as that of \citet{tt_2016} and \citet{tmt_2020}, who 
used it to study nascent planets in viscous discs;
we will explore both viscous and inviscid discs.

The problem of planetary accretion within disc gaps
is also a global one insofar as a planet
can 
accrete 
gas that is brought to it from afar,
from regions outside the gap. 
Thus we will stage our calculations
within circumstellar discs that transport
mass across decades in radius. This opens up another
form of feedback: in feeding the planet, the disc 
can have its entire surface density profile changed 
(e.g., \citealt{ld_2006}; \citealt{znh_2011}; \citealt{o_2016}).

Our work is organized as follows. In section 
\ref{sec:consumption} we describe how Lindblad repulsion
and planetary accretion of disc gas (``consumption'') work
together to determine gap depths and
surface density profiles of viscous circumstellar
accretion discs. 
Our largely analytic considerations are supplemented
with simple numerical experiments modeling planet-disc
interactions and disc evolution in 1D (orbital radius).
In section \ref{sec:consumption} we fix,
for simplicity, the planet mass; in section \ref{sec:mass},
we allow the planet mass to grow freely and solve
the full two-way feedback problem. In section \ref{sec:windy},
motivated by recent theoretical and observational developments,
we consider discs that transport their mass not by viscous
diffusion but rather by angular momentum losses from magnetized
winds. For such inviscid, wind-driven discs, accretion is not diffusive
but purely advective, and embedded planets carve out especially
deep gaps in the absence of viscous backflow.
We summarize and discuss the implications of our findings
on gas giant masses and disc structure, including the
structure of transitional discs, in section \ref{sec:sum}.

A simplified study
such as ours will not capture important (and sometimes
poorly understood) effects, among them planetary migration
(e.g., \citealt{kn_2012}; \citealt{dhm_2014}; \citealt{dk_2015,dk_2017}; \citealt{fc_2017}; \citealt{kts_2018}; \citealt{mnp_2020}),
eccentricity evolution (both of the planet and the disc;
e.g., \citealt{pnm_2001}; \citealt{gs_2003}; \citealt{kd_2006}; \citealt{dc_2015}; \citealt{mfv_2019}), and
the 3D dynamics of circumplanetary discs (e.g., \citealt{fzc_2019}).
Our goal is not so much to be realistic but to acquire some intuition
about the interplay of Lindblad repulsion and planetary accretion,
and to provide
a baseline understanding that can guide the development and interpretation
of more sophisticated models. Where possible, we place our results in context
with state-of-the-art numerical experiments in the literature (see in particular section \ref{sec:sum}).

\section{Viscous discs: Surface Density Profiles at Fixed Planet Mass} \label{sec:consumption}

We study how
the surface densities of viscous accretion
discs are shaped by repulsive planetary Lindblad torques in addition
to planetary accretion of disc gas (``consumption'').
Section \ref{sec:oom} contains analytic considerations which
are tested 
numerically
in section \ref{sec:num}.
In these sections, while we allow the disc surface density
to deplete by consumption, we do not simultaneously
allow the planet's mass to increase. This fixing of the planet's
mass is done for simplicity,
to see how the planet affects the disc
but not vice versa.
In section \ref{sec:mass},
we free up the planet's mass and allow two-way feedback
between planet and disc.

\subsection{Order-of-magnitude scalings}\label{sec:oom}

\begin{figure}
    \centering
    \includegraphics[width=\linewidth]{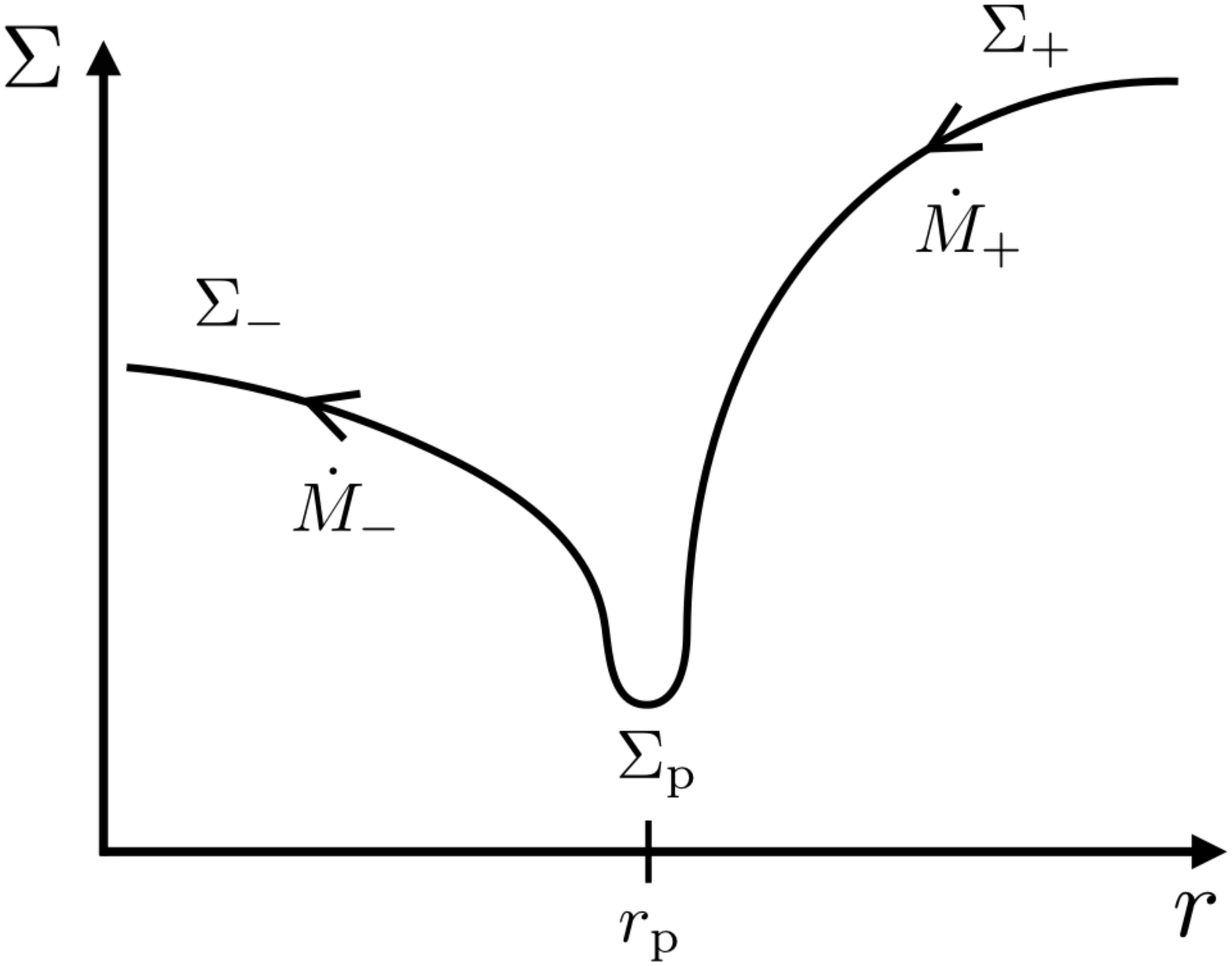}
    \caption{Sketch of the disc surface density and  accretion flow in the vicinity of a planet. The planet is located at orbital radius $r_\mathrm{p}$, inside a gap having surface density $\Sigma_\mathrm{p}$. At $r > r_{\rm p}$, the disc surface density is $\Sigma_+$ and mass accretes inward at rate $\dot{M}_+$. Downstream of the planet, at $r < r_{\rm p}$, the corresponding surface density and accretion
    rate are $\Sigma_-$ and $\dot{M}_-$, respectively. The difference $\dot{M}_+ - \dot{M}_-$ is the accretion
    rate onto the planet $\dot{M}_{\rm p}$.}
    \label{fig:sig_cartoon}
\end{figure}

Consider an accreting planet embedded in a viscous disc.
From Figure \ref{fig:sig_cartoon} we identify three disc surface densities: 
$\Sigma_{\rm p}$ at the orbital radius of the
planet ($r = r_{\rm p}$),
$\Sigma_+$ exterior to the planet, and $\Sigma_-$
interior to the planet. The planet depresses
the local surface density because it is both consuming
disc gas and repelling disc gas away by
Lindblad torques. 
Our goal is to estimate the depth of the planet's gap 
in relation to the inner and outer discs:
$\Sigma_{\rm p}/\Sigma_-$ and $\Sigma_{\rm p}/\Sigma_+$.
We assume a steady state where the disc
has viscously relaxed: given a viscosity $\nu$, the system age $t$ is at least as long as the diffusion time $r^2/\nu$
across the disc. In addition, $t$ is at most the planet growth
timescale $M_{\rm p}/\dot{M}_{\rm p}$, so that we
may consider the planet mass fixed
at any given moment.

Mass flows steadily
inward at rate $\dot{M}_+$ from the outer disc. 
Part of this flow is accreted by the planet at rate
$\dot{M}_{\rm p}$, with the rest feeding
the inner disc which accretes onto the star at rate $\dot{M}_-$. 
Dropping numerical pre-factors (these will be
restored in later sections), we have
\begin{align}
\dot{M}_+ &= \dot{M}_- + \dot{M}_{\rm p} \nonumber \\
\Sigma_+ \nu &\sim \Sigma_- \nu + \dot{M}_{\rm p} \nonumber \\
&\sim \Sigma_- \nu + A \Sigma_{\rm p} \,. \label{eqn:mass}
\end{align}
There are a number of assumptions embedded in these
order-of-magnitude statements.
For $\dot{M}_+$ and $\dot{M}_-$ we have substituted
standard steady-state expressions
for a disc of shear viscosity $\nu$ (e.g., \citealt{fkr_2002}), valid asymptotically at locations far
from any mass sink  ($|r-r_{\rm p}| \gtrsim r_{\rm p}$).
At the same time, the locations we are considering
in the outer and inner discs are not so far 
from the planet that we need to account for spatial variations
in $\nu$, which may change by order-unity factors over
length scale $r$.

For the planet's accretion rate, we have assumed in
(\ref{eqn:mass}) that it
scales linearly with the local surface density $\Sigma_{\rm p}$
with proportionality constant $A$:
\begin{equation}\label{eqn:Adef}
\dot{M}_{\rm p} = A \Sigma_{\rm p} \,.
\end{equation}
This assumption is satisfied, e.g., by a planet
accreting at the Bondi rate (e.g., \citealt{fkr_2002}):
\begin{align} \label{eq:m_dot_bondi_simp}
    \dot{M}_{\rm p, Bondi} &\sim \rho_{\rm p} c_{\rm s} R_{\rm B}^2 \nonumber \\
    &\sim \frac{\Sigma_{\rm p}}{H} c_{\rm s} \left( \frac{GM_{\rm p}}{c_{\rm s}^2} \right)^2  
\end{align}
where $\rho_{\rm p}$ is the disc midplane mass
density near the planet, $c_{\rm s}$ is the disc sound speed,
$R_{\rm B} = GM_{\rm p}/c_{\rm s}^2$ is the Bondi radius,
$H = c_{\rm s}/\Omega$ is the disc scale height,
$\Omega$ is the orbital angular frequency, 
and $G$ is the gravitational constant. Then
\begin{equation} \label{eqn:A}
A_{\rm Bondi} \sim \frac{m^2}{h^4} \Omega r^2
\end{equation}
where $m \equiv M_{\rm p}/M_\star$ is the planet-to-star
mass ratio, and $h\equiv H/r$ is the disc aspect ratio.
Ginzburg \& Chiang (\citeyear{gc_2019a}, their section 1.1) discusses how Bondi
accretion may be valid for ``sub-thermal'' planets whose
masses are less than 
\begin{equation} \label{eqn:thermalmass_0}
M_{\rm thermal} \sim h^3 M_\star 
\end{equation}
the mass for which the Bondi radius $R_{\rm B}$, the
Hill radius $R_{\rm H} \sim m^{1/3} r$, and the
disc scale height $H$ are all equal. 
A sub-thermal planet has
$R_{\rm B} < R_{\rm H} < H$---its
gravitational sphere of influence has radius $R_{\rm B}$, set
by gravity and thermal pressure---and should accrete at the Bondi rate,
isotropically from the all-surrounding disc (\citealt{gc_2019a}; 
see also fig.~1 of \citealt{tt_2016} for 
evidence supporting the Bondi $m^2$ scaling,
taken from the 3D simulations of
\citealt{dkh_2003}). For a super-thermal planet
having $M>M_{\rm thermal}$, the hierarchy of length scales
reverses so that 
$R_{\rm B} > R_{\rm H} > H$---the planet's sphere of influence, now set by gravitational tides at radius $R_{\rm H}$, ``pops out'' of the disc---and 
arguably 
the planet accretes in a more 2D fashion, presenting a cross-section of order $R_{\rm H} H$ to disc gas
that shears by at a velocity $\Omega R_{\rm H}$.
The corresponding ``Hill rate'' for consumption is then
\begin{equation} \label{eqn:hillsimple}
\dot{M}_{\rm p,Hill} \sim \rho_{\rm p} \times R_{\rm H}H \times \Omega R_{\rm H} \sim \Sigma_{\rm p} R_{\rm H}^2 \Omega
\end{equation}
whence
\begin{equation}
A_{\rm Hill} \sim m^{2/3} \Omega r^2 \,.
\end{equation}
A Hill-based scaling for consumption is commonly used
in 2D disc-planet hydrodynamical simulations (e.g., \citealt{znh_2011,dk_2015,dk_2017,mfv_2019}).
We have assumed in writing the above that the planet masses are large enough
for accretion to be hydrodynamically-limited
as opposed to cooling-limited (\citealt{gc_2019a},
cf.~their fig.~1).

In this paper we will calculate the growth of planets
from sub-thermal to super-thermal masses, so will have occasion to
use both $A_{\rm Bondi}$ and $A_{\rm Hill}$.
We recognize that the 2D picture motivating our
Hill scaling may not be correct; in 3D, meridional flows
from gap walls can feed the planet along its poles
\citep{smc_2014,msc_2014,fc_2016}. Relatedly, the disc density 
scales with height $z$ above the midplane 
as $\exp[-z^2/(2H^2)]$ (for an isothermal atmosphere),
which implies that a considerable fraction of the 
disc mass resides between $|z|= H$ and $2H$; accordingly, the planet does not
pop out of the disc until it is strongly
super-thermal, i.e., until $m$ is a large
multiple of $h^3$ (cf.~equation \ref{eqn:thermalmass_0}). 
An isotropic
version of super-thermal accretion 
controlled by the Hill sphere gives 
$\dot{M}_{\rm p,Hill,iso} \sim \rho_{\rm p} \times R_{\rm H}^2 \times \Omega R_{\rm H}$
or $A_{\rm Hill,iso} \sim m \Omega r^2 / h$.
Yet another prescription for
accretion is given by 
\citet{tw_2002}: $A_{\rm TW} \sim m^{4/3} \Omega r^2 / h^2$, an empirical relation based on their
2D numerical simulations (see also \citealt{tt_2016}).
To the extent that these alternative
scalings increase with $m$ more steeply 
than our nominal
$A_{\rm Hill} \propto m^{2/3}$,
whatever final super-thermal planet masses we derive 
should be lower limits 
(see sections \ref{sec:proc} and \ref{sec:sum}).


Momentum conservation provides another relation between
the surface densities. It is easiest to write down
downstream of the planet in the accretion flow (in the inner disc),
as the flow of momentum upstream (in the outer disc)
is complicated by the mass sink presented by the planet.
In the inner disc there are no sinks of mass or momentum,
only a steady transmission of mass inward 
and angular momentum outward (assuming, as we do throughout
this paper, a non-migrating planet; see section \ref{sec:sum}
for pointers to the migrating case).
The rate at which
angular momentum is carried viscously outward by
the inner disc equals the viscous transport rate 
local to the planet, plus the repulsive
Lindblad torque exerted
by the planet on the disc:\footnote{The planet
excites waves in the inner disc which carry negative
angular momentum inward. This is equivalent to transmitting
positive angular momentum outward.}
\begin{align}
\Sigma_- \nu \Omega r^2 \sim \Sigma_{\rm p} \nu \Omega r^2 + B \Sigma_{\rm p} \Omega r^2 
\label{eqn:momentum}
\end{align}
with 
\begin{equation} \label{eqn:B}
B \sim \frac{m^2}{h^3} \Omega r^2 
\end{equation}
given by the standard \citet{gt_1980}
linear Lindblad torque, 
integrating the effects of all Lindblad
resonances up to the torque cutoff.
A similar statement to (\ref{eqn:momentum}), dropping the
viscous term local to the planet, was made by \citet{fsc_2014}. Given $B$, (\ref{eqn:momentum})
can be solved for the gap contrast
with the inner disc:
\begin{align}
\frac{\Sigma_{\rm p}}{\Sigma_-} \sim \frac{1}{1 + B/\nu} 
\label{eqn:oom1}
\end{align}
(see also \citealt{dm_2013}; \citealt{kmt_2015}; \citealt{gs_2018}). 
Combining mass conservation (\ref{eqn:mass}) with
momentum conservation (\ref{eqn:momentum})
yields the gap contrast with the outer disc:
\begin{align}
\frac{\Sigma_{\rm p}}{\Sigma_+} \sim \frac{1}{1 + (A + B)/\nu} \,.
\label{eqn:oom2}
\end{align}
An equivalent equation 
is derived by
\citet[][their appendix B]{tt_2016} and  \citet[][their equation 26]{tmt_2020}. Equations (\ref{eqn:oom1}) and (\ref{eqn:oom2}) inform
us that planetary consumption ($A \neq 0$) leads
to asymmetric gap contrasts: a deeper gap
relative to the outer disc than to the inner disc. 
The outer gap contrast is the more important insofar
as the outer disc controls surface densities
everywhere downstream; in other words,
$\Sigma_+$ is the independent variable
while $\Sigma_{\rm p}$ and $\Sigma_-$ are dependent variables.
Equation (\ref{eqn:oom2}) states that, given $\Sigma_+$, 
the effects of accretion ($A$) and repulsion ($B$) 
in setting the gap depth $\Sigma_{\rm p}$ 
are additive (not multiplicative). If $A > B$, then 
consumption dominates.

For $A=A_{\rm Bondi}$ and $B$ given by (\ref{eqn:B}),
\begin{equation}\label{eqn:ab}
A_{\rm Bondi}/B \sim 1/h > 1
\end{equation}
and consumption dominates repulsion in setting the gap depth,
independent of planet mass in the sub-thermal regime.
On the other hand, for $A=A_{\rm Hill}$,
\begin{equation}\label{eqn:ab_Hill}
A_{\rm Hill}/B \sim m^{-4/3} h^3 
\end{equation}
which says that for super-thermal planets that are massive
enough,
repulsion dominates consumption ($A_{\rm Hill}/B < 1$).

We may also solve for the relative accretion rates:
\begin{align}
\frac{\dot{M}_{\rm p}}{\dot{M}_+} &\sim \frac{A\Sigma_{\rm p}}{\Sigma_+\nu} \sim  \frac{A/\nu}{1+(A+B)/\nu} \label{eq:m_dot_ratio} \\
 \frac{\dot{M}_-}{\dot{M}_+} &\sim \frac{\Sigma_-}{\Sigma_+} \sim \frac{1+B/\nu}{1+(A+B)/\nu} \,. \label{eqn:mdot-0}
\end{align} 
A couple example limiting cases of (\ref{eq:m_dot_ratio})
and (\ref{eqn:mdot-0}) are as follows.
If we take $A/B = A_{\rm Bondi}/B \sim 1/h > 1$ 
and further assume that $B/\nu > 1$ so that the inner gap contrast
is significant (equation \ref{eqn:oom1}), we find
\begin{align}
    \frac{\dot{M}_{\rm p}}{\dot{M}_+} &\sim 1 - B/A_{\rm Bondi} \sim 1 - h \label{eqn:mdotp}\\
    \frac{\dot{M}_-}{\dot{M}_+} &\sim B/A_{\rm Bondi} \sim h \label{eqn:mdot-}
\end{align}
which says that the planet consumes
nearly all of the mass supplied to it by
the outer disc, leaving behind a fraction 
$h$ to feed the inner disc.
If instead we take $A/B = A_{\rm Hill}/B$
and further assume 
$B > A_{\rm Hill} > \nu$ (repulsion-limited and deep gap), then
\begin{align}
    \frac{\dot{M}_{\rm p}}{\dot{M}_+} &\sim A_{\rm Hill}/B \sim m^{-4/3} h^3 < 1 \label{eqn:mdotp_Hill}\\
    \frac{\dot{M}_-}{\dot{M}_+} &\sim 1 - A_{\rm Hill}/B \sim 1 - m^{-4/3} h^3 \label{eqn:mdot-_Hill}
\end{align}
and the planet diverts only a small fraction, $A_{\rm Hill}/B$,
of the disc accretion flow onto itself.

The order-of-magnitude considerations presented here
are firmed up 
in subsequent sections, including in Appendix \ref{sec:analy_ss}, where we
derive in greater analytic detail the surface density profile
and mass accretion rates, drawing from \citet{ld_2006}.

\subsection{Numerical simulations}\label{sec:num}

\subsubsection{Procedure}\label{sec:proc}

We solve numerically for the 1D evolution
of a viscously shearing disc \citep[e.g.,][]{fkr_2002}
with a planetary mass sink.
The governing equation for the surface density
$\Sigma(r,t)$ in cylindrical radius $r$ and time $t$ reads 
\begin{align}
\pd{\Sigma}t=\frac{3}{r}\pd{}r\left[r^{1/2}\pd{}r\left(r^{1/2}\nu\Sigma\right)\right] - \frac{\dot{M}_{\rm p}(t)}{2\pi r} \delta (r - r_{\rm p}) 
\label{eq:dsig_dt}
\end{align} 
where $\delta$ is the Dirac delta function 
and $r_{\rm p}$ is the radial position of the planet
(held fixed). For the viscosity $\nu$ we employ
the \citet{ss_alpha} $\alpha$-prescription:
\begin{equation} \label{eq:ss_alpha}
\nu = \alpha c_{\rm s}^2/\Omega = \alpha h^2 \Omega r^2
\end{equation}
where $\Omega$ is the Keplerian orbital
frequency around a $1 \, M_\odot$ star,
$c_{\rm s} = \sqrt{k_{\rm B}T/\overline{m}}$, 
the disc temperature is $T = 200 \, {\rm K} (r/{\rm au})^{-1/2}$,
$k_{\rm B}$ is Boltzmann's constant, $\overline{m} = 2m_{\rm H}$ is the mean molecular mass, $m_{\rm H}$ is the mass of the hydrogen atom, and
\begin{equation}
h \equiv H/r = c_{\rm s}/(\Omega r) \simeq 0.054 \left( \frac{r}{{\rm 10 \, au}} \right)^{1/4} \,.
\end{equation}
We fix $\alpha = 10^{-3}$ for the results in this section. Given these inputs,
$\nu = \nu(r) \propto r^1$.

Apart from the mass sink, equation (\ref{eq:dsig_dt}),
which combines the 1D mass and momentum equations, is
identical to the diffusion equation governing
an isolated viscous disc as derived by \citet{lp_1974}.
What is missing 
is an explicit accounting for the repulsive
Lindblad torque exerted by the planet. Many studies
include the planetary torque by introducing, into the momentum
equation, a term for the torque per unit radius that scales as
${\rm sgn}(x)/x^4$, where
$x \equiv r - r_{\rm p}$
(e.g., \citealt{lp_1986}; \citealt{ld_2006}).
Compared against 2D hydrodynamical simulations,
this $1/x^4$ prescription has been shown in 1D studies
to reproduce the azimuthally averaged surface density
profiles of repulsive gaps near their peripheries (at $x \gtrsim 4 H$) but not near gap centers (at $x \lesssim 4 H$;
\citealt{fsc_2014}, their section 4.3).
In particular the 1D torque density prescription,
which assumes angular momentum is deposited
locally and neglects wave propagation, fails to recover the flat bottoms
of gaps and the surface densities there (cf.~\citealt{gs_2018} who use the \citealt{gr_2001} wave steepening theory to lift these assumptions). 
This shortcoming of the $1/x^4$ prescription means that it cannot be used to compute the planetary accretion
rate $\dot{M}_{\rm p}$, which depends
on knowing the gas density in the planet's immediate
vicinity.

What we do instead to include the repulsive Lindblad torque
when calculating planetary accretion is as follows.
Within the radial grid cell at $r = r_{\rm p}$
of width $\Delta r_{\rm p}$, the
surface density is reduced after every timestep $\Delta t$
according to 
\begin{align} \label{eqn:euler}
\Sigma (r_{\rm p}, t+\Delta t) = \Sigma (r_{\rm p}, t) - \frac{\dot{M}_{\rm p} (t) \Delta t}{2 \pi r_{\rm p} \Delta r_{\rm p}} \,\,\,\,\,\,\,\,\,\,\,\,\,\,\,\,\,\,\,\,\,\, ({\rm simulation})
\end{align}
where the label ``simulation'' reminds us that this equation
applies to the numerical simulation only
and should not be used outside of that context.
It is in evaluating $\dot{M}_{\rm p}$ that
we include, in a ``sub-grid'' manner,
the repulsive Lindblad gap: 
\begin{align} \label{eqn:subgrid}
\dot{M}_{\rm p}(t) &= A \times \frac{\Sigma (r_{\rm p},t)}{1+B/\nu} \,\,\,\,\,\,\,\,\,\,\,\,\,\, \, \, \,\,\,\,\,\,\,\,\,\,\,\,\,\, \, \,\,\,\,\,\,\,\,\,\,\,\,\,\, \,\,\,\,\,\,\,({\rm simulation}).
\end{align}
What equation (\ref{eqn:subgrid}) says is that
the disc surface density the planet actually ``sees''
when consuming local gas 
is lower than the numerically computed
``grid-level'' surface density $\Sigma (r_{\rm p},t)$---lower
by the Lindblad reduction factor $1/(1+B/\nu)$
(equation \ref{eqn:oom1}). In other words,  
repulsion is encoded/enforced at a sub-grid level.
We stress that equation (\ref{eqn:subgrid}) is used
only in our numerical simulation to capture repulsion
and should not be used outside of it; contrast
(\ref{eqn:subgrid}) with, e.g., (\ref{eq:m_dot_ratio}),
and note that $\Sigma(r_{\rm p},t)$ is notation
specific to the simulation and should not be confused with $\Sigma_{\rm p}$, the
actual surface density at the planet's position.

Our numerical procedure captures the gap depth but not the gap width,
as the sub-grid modification is restricted (for simplicity)
to the grid cell containing the planet. 
We consider this crude scheme acceptable
insofar as we are more interested in the
gross magnitudes for $\Sigma_{\rm p}/\Sigma_+$
and $\Sigma_{\rm p}/\Sigma_-$ and less interested 
in the precise surface density gradients.
An untested assumption underlying
our numerical procedure---and in our
steady-state analytics---is
that material flows radially through
the gap at whatever velocity $u_r$ is 
needed to maintain continuity,
i.e., to enforce
$\dot{M}_- = \dot{M}_+ - \dot{M}_{\rm p} = -2\pi \Sigma_{\rm p} r_{\rm p} u_r$ (where $u_r < 0$ for accretion toward the star).
We cannot test this assumption as we do not resolve the flow dynamics inside the gap. We will call out this
assumption in the results to follow (sections \ref{sec:nc10au} and \ref{sec:numwind}).
Also, as a reminder, we note that while the surface density 
changes as a result of consumption, 
in this subsection we fix $M_{\rm p}$,
i.e., we do not update $M_{\rm p}$ using $\dot{M}_{\rm p}$ 
(this assumption is relaxed in section \ref{sec:mass}).

In evaluating the consumption and repulsion coefficients $A$ and $B$,
we make choices similar to those in our earlier
order-of-magnitude analysis (section \ref{sec:oom}), 
except that now we include numerical pre-factors for
greater precision:
\begin{align}
A_{\rm Bondi} &= 
0.5 \, \Omega r^2 \frac{m^2}{h^4}  && {\rm for} \, \subthermal \,\, m \leq 3 h^3  \label{eqn:bondi}\\
A_{\rm Hill} &= 2.2 \, \Omega r^2 m^{2/3} && {\rm for} \, \superthermal \,\, m > 3 h^3 \label{eqn:hill}\\
B &= 0.04 \, \Omega r^2 \frac{m^2}{h^3}
\label{eqn:kanagawa}
\end{align}
where all quantities are evaluated at $r_{\rm p}$.
The pre-factor of $0.5$
in equation (\ref{eqn:bondi})
is calibrated using 3D simulation results
for $\dot{M}_{\rm p}$ from D'Angelo et al.~(\citeyear{dkh_2003}; these are re-printed in fig.~1 of \citealt{tt_2016}). 
The coefficient of 2.2 in equation (\ref{eqn:hill})
follows from requiring that (\ref{eqn:bondi}) match (\ref{eqn:hill})
at the thermal mass
\begin{equation} \label{eqn:thermalmass}
M_{\rm thermal} \equiv 3 h^3 M_\star \simeq 0.5 \left( \frac{h}{0.054} \right)^3 M_{\rm J}
\end{equation}
defined by equating $H$ with
$R_{\rm H} = (m/3)^{1/3} r$, 
with $M_{\rm J}$ the mass of Jupiter. 
Equation (\ref{eqn:kanagawa}) is taken from
the numerical 2D simulations of \citet[][see also \citealt{dm_2013} and \citealt{d_2015}
who report similar results]{kmt_2015}. 

Note further that the expressions we used in section
\ref{sec:oom} for the steady disc
accretion rates $\dot{M}_+$ and $\dot{M}_-$ should
be amended with the numerical pre-factor $3\pi$, i.e.,
$\dot{M}_+ = 3\pi \Sigma_+ \nu$ and similarly for
$\dot{M}_-$ (e.g., \citealt{fkr_2002}). 
This correction is already embedded in the
diffusion equation (\ref{eq:dsig_dt}). 
Including this pre-factor in equation (\ref{eqn:mass})
implies that $A$ should be replaced with $A/(3\pi)$
in equations (\ref{eqn:oom2})--(\ref{eqn:mdot-_Hill}).
Putting it all together, we have
\begin{align} \label{eq:A_B_visc}
    \frac{A_{\rm Bondi}}{3 \pi B} \simeq \frac{1.3}{h} > 1
\end{align}
implying that consumption always dominates for sub-thermal masses. Furthermore, 
\begin{align} \label{eq:A_B_visc_Hill}
    \frac{A_{\rm Hill}}{3 \pi B} \simeq 5.5 m^{-4/3} h^3 \simeq 1.0 \left( \frac{m}{5 \times 10^{-3}} \right)^{-4/3} \left( \frac{h}{0.054} \right)^3
\end{align}
implying that repulsion dominates for super-thermal masses exceeding a ``repulsion mass'' 

\begin{align} \label{eqn:repulsionmass}
M_{\rm repulsion,visc} &\simeq 3.6 h^{9/4} M_\star \nonumber \\
& \simeq 5.3 M_{\rm J} \left( \frac{h}{0.054} \right)^{9/4} \simeq 5.3 M_{\rm J} \left( \frac{r}{10 \, {\rm au} } \right)^{9/16} \nonumber \\
&\simeq 1.2 h^{-3/4} M_{\rm thermal} \simeq 11 \left( \frac{0.054}{h} \right)^{3/4} M_{\rm thermal}  \,.
\end{align}
For $M<M_{\rm repulsion,visc}$, consumption dominates and the planet
accretes nearly all the disc gas that tries to diffuse past;
for $M > M_{\rm repulsion,visc}$, repulsion dominates and the planet's accretion rate falls below the
disc accretion rate. 
The above expression for $M_{\rm repulsion, visc}$ depends on our
assumption that planetary accretion follows our Hill
scaling $A_{\rm Hill} \propto m^{2/3}$ for super-thermal masses. As discussed in
section \ref{sec:oom}, this assumption might not be correct. 
If instead of $A_{\rm Hill}$
we use $A_{\rm TW} = 0.29 \,\Omega r^2 \,m^{4/3}/h^2$
as found from the 2D numerical simulations of
\citet{tw_2002}, we would 
find $A_{\rm TW}/(3\pi B) \simeq 4 \,( M_\mathrm{p}/M_\mathrm{J} )^{-2/3} ( h/0.054)$, in which case the mass above which repulsion dominates would change to
$M_{\rm repulsion,visc,TW} \simeq 9 \, M_{\rm J} \,[r / (10 \, {\rm au})]^{3/8}$. This is nearly twice the value of $M_{\rm repulsion,visc}$ given by (\ref{eqn:repulsionmass}), and would imply a more extended consumption-dominated growth phase. 
Insofar as our nominal model
adopts $A_{\rm Hill}$ which leads to a more limited
consumption-dominated growth phase, the planet masses
we compute for our viscous disc model are lower limits.

So far we have described how we compute
the mass sink term, which includes the sub-grid
Lindblad torque, in equation (\ref{eq:dsig_dt}).
The remaining diffusive term is solved in a standard
way. We first change variables to 
$z \equiv r^{1/2} \nu \Sigma$ and $y \equiv 2 r^{1/2}$
so that the diffusive portion of 
equation \eqref{eq:dsig_dt} reads
\begin{align} \label{eq:kep_simp}
\pd{z}t &= \frac{12 \nu}{y^2} \pd{^2 z}{y^2}
\end{align}
with non-constant diffusion coefficient $12 \nu/y^2$.
We solve equation \eqref{eq:kep_simp} 
as an initial value problem 
using an implicit scheme (e.g., \citealt{ptv_2007}).
Our computation grid extends from an inner boundary of $r_\mathrm{in} = 0.01 \, \mathrm{au}$ to an outer boundary of $r_\mathrm{out} = 500 \, \mathrm{au}$,
and is divided into 300 cells that are uniform in $\Delta y$.
We fix the timestep $\Delta t = 10^{-4} t_{\nu,{\rm p}}$, 
where $t_{\nu,{\rm p}} \equiv r_{\rm p}^2 / \nu (r_{\rm p}) \simeq 1.7$ Myr is the viscous diffusion timescale
at the planet's orbital radius of $r_{\rm p} = 10$ au (where $h \simeq 0.054$).
Recognizing that our transformed variable $z$ is
proportional to the viscous torque $2 \pi \nu \Sigma r^3  d \Omega / dr \propto r^{1/2} \nu \Sigma$,
we use a torque-free inner boundary condition,
$z (r_\mathrm{in}) = 0$,
as would be the case if the disc were truncated
by a co-rotating stellar magnetosphere 
(shearless boundary layer).
At the outer boundary
we assume the torque gradient 
$\partial z / \partial r \,(r_{\rm out}) = 0$. Neither
boundary condition is critical as we are
interested in the flow near the planet,
away from either boundary.

The surface density of the disc is initialized with
the similarity solution for an isolated viscous accretion disc 
with $\nu \propto r^1$
(\citealt{lp_1974}; \citealt{hcg_1998}):
\begin{align} \label{eqn:sim}
    \Sigma(r,0) = \frac{M_{\rm disc}}{2 \pi r_1^2} \frac{r_1}{r}  e^{-r/r_1} \,\,\,\,\,\,\,\,\, \,\,\,\,\,\,\,\,\, \,\,\,\,\,\,\,\,\, \,\,\,\,\,\,\,\,\, \,\,\,\,\,\,\,\,\, 
    ({\rm simulation})
\end{align}
where $M_{\rm disc}  = 15.5 \, M_{\rm J} \simeq 0.015  \, M_\odot$
is the initial mass of the disc 
and $r_1 = 30$ au is a characteristic disc radius 
(where the diffusion time is $r_1^2/\nu \simeq 5$ Myr). 
We consider two fixed planet masses,
$M_{\rm p} = 0.3 \,M_{\rm J} < M_{\rm thermal}$ 
and $M_{\rm p} = 10 \,M_{\rm J} > M_{\rm thermal}$. 
Planet masses that freely grow
are modeled in section \ref{sec:mass}.

At every timestep, we first advance
$\Sigma(r,t)\rightarrow \Sigma(r,t+\Delta t)$
for all $r$ according to (\ref{eq:kep_simp})
using the implicit solver, and then
we advance $\Sigma(r_{\rm p},t)\rightarrow \Sigma(r_{\rm p},t+\Delta t)$ using (\ref{eqn:euler})
and (\ref{eqn:subgrid}).
This procedure is repeated 
until the disc is evolved for
several $t_{\nu,{\rm p}}$,
long enough for the disc near the planet to achieve 
a quasi-steady state.

\begin{figure}
    \centering
    \includegraphics[width=\linewidth]{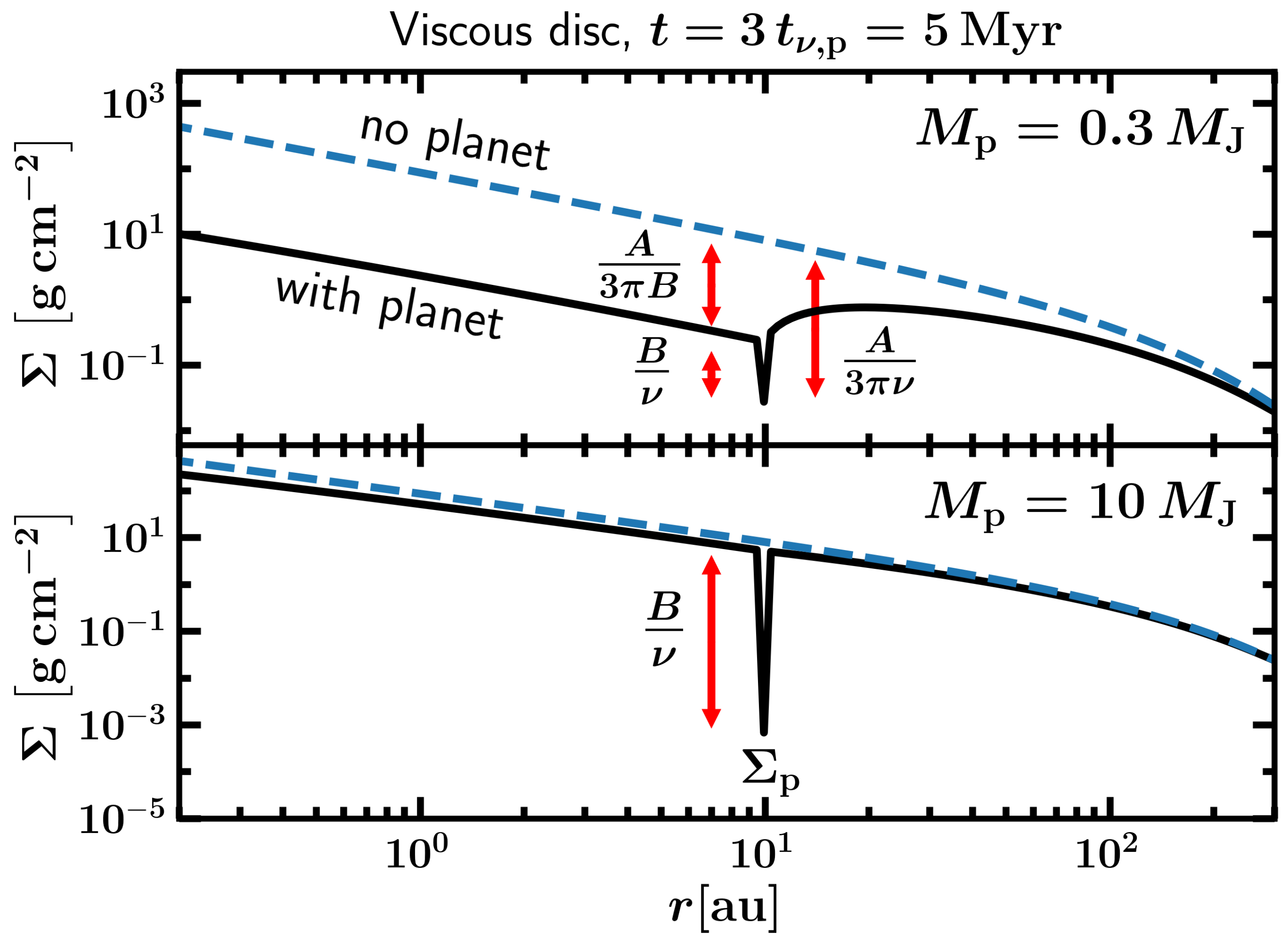}
    \caption{How the surface density profile of a viscous disc responds to a planet that both consumes disc gas, 
    and repels gas away by Lindblad torques. Surface densities are calculated from our 1D numerical simulation of a planet of fixed mass, either $M_{\rm p} = 0.3 \,M_{\rm J}$ (top panel) or $M_{\rm p} = 10\, M_{\rm J}$ (bottom panel),
    at $t = 3 t_{\nu,\mathrm{p}}$ when the disc near the planet at $r_{\rm p} = 10$ au has viscously relaxed.
    When computing
    the planetary accretion rate $\dot{M}_{\rm p}$,
    the gap is modeled as a single cell whose ``true''
    surface density equals the grid-level
    $\Sigma$ lowered by a factor of $(1+B/\nu) \simeq B/\nu$; plotted here are the true sub-grid values $\Sigma_{\rm p}$. 
    Accordingly, the planet's gap is not
    spatially resolved and its width should not be taken
    literally from this figure. 
    Red double-tipped arrows have lengths equal to their associated variables in dex, and demonstrate
    good agreement between numerics and analytics.
    The planet of mass $M_{\rm p} = 0.3 \,M_{\rm J}$,  accreting at the Bondi rate, creates an asymmetric gap,
    with the inner disc surface density $\Sigma_-$ lower than the outer $\Sigma_+$ by $A_{\rm Bondi}/(3\pi B) > 1$;
    conditions are always consumption-dominated for
    Bondi accretion and $B$ as given by (\ref{eqn:kanagawa}). The planet
    of mass $M_{\rm p} = 10\, M_{\rm J}$, accreting at
    the Hill rate, creates a symmetric gap where $\Sigma_-/\Sigma_+ \sim 1$; conditions here are repulsion-dominated as $M_{\rm p} > M_{\rm repulsion,visc}$ (equation \ref{eqn:repulsionmass}).}
    \label{fig:sigma_detailed}
\end{figure}

\subsubsection{Results}\label{sec:res222}

Figure \ref{fig:sigma_detailed} shows, for $M_{\rm p} = \{0.3, 10\} M_{\rm J}$, the numerically computed 
surface density profiles $\Sigma(r)$ 
at $t = 3 t_{\nu,{\rm p}} \simeq 5$ Myr.
Overlaid for comparison is our numerical solution without a planet,
which we have verified matches the analytic
time-dependent similarity solution of \citet{lp_1974}.
For the case with a planet, rather
than plot at face value
the numerically computed (grid-level) $\Sigma (r_{\rm p},t)$, 
we plot that value multiplied by the sub-grid reduction
factor $1/(1+B/\nu)$---this is the ``true'' value
for $\Sigma_{\rm p}$ that incorporates
the repulsive Lindblad torque. 
Since this sub-grid correction factor is applied to only
a single grid point, we cannot resolve
gap widths; our focus instead is on the gross gap
contrasts $\Sigma_{\rm p}/\Sigma_+$
and $\Sigma_{\rm p}/\Sigma_-$.

The surface density profiles shown
in Figure \ref{fig:sigma_detailed}
conform to the analytic considerations
of section \ref{sec:oom}.
For $M_{\rm p} = 0.3 \, M_{\rm J}$ (top panel),
conditions are consumption-limited: 
$\Sigma_+/\Sigma_{\rm p} \sim A_{\rm Bondi}/(3\pi\nu)$
(equation \ref{eqn:oom2}
in the limit $A_{\rm Bondi}/(3\pi) > B > \nu$)
and the surface density of the entire interior
disc is depressed relative to the same disc
without a planet by a factor of
$\Sigma_+ / \Sigma_- \sim \dot{M}_+ / \dot{M}_- \simeq A_{\rm Bondi}/(3\pi B)$
(equations \ref{eqn:mdot-0} and \ref{eqn:mdot-}).
By comparison, for $M_{\rm p} = 10 \, M_{\rm J}$ (bottom panel),
the gap is more nearly symmetric, $\Sigma_+/\Sigma_- \sim 1$
(equations \ref{eqn:mdot-0} and \ref{eqn:mdot-_Hill}), and
deep and repulsion-dominated,
$\Sigma_+/\Sigma_{\rm p} \sim B/\nu$ (equation \ref{eqn:oom2}
in the limit $B > A_{\rm Hill}/(3\pi) > \nu$).

\begin{figure}
    \centering
    \includegraphics[width=\linewidth]{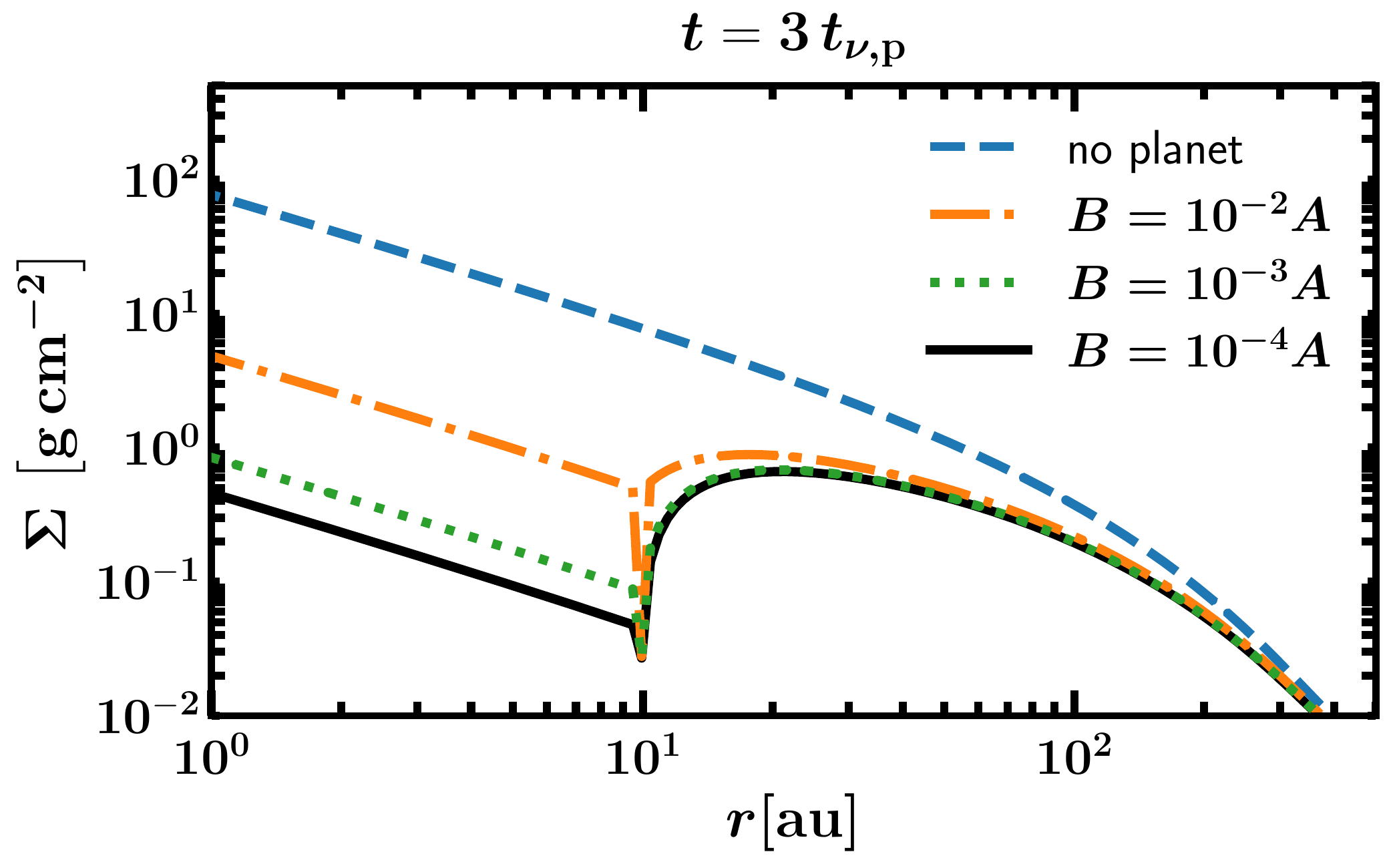}
    \caption{Same as Figure \ref{fig:sigma_detailed} for the
    case $M_{\rm p} = 0.3 M_{\rm J}$, but for different
    choices of $B$ scaled to $A_{\rm Bondi}$. As long as $A_{\rm Bondi}/(3 \pi) > B$, the planet's gap is consumption-dominated and its surface density $\Sigma_\mathrm{p}$ is independent
    of the repulsion coefficient $B$. The depression of the inner disc relative to the outer disc is, however,
    sensitive to $B$ for $B > \nu$;
    $\Sigma_-/\Sigma_+ \simeq (1+B/\nu)/[A_{\rm Bondi}/(3\pi\nu)]$.
    }
    \label{fig:m_final_visc}
\end{figure}

So long as consumption is stronger than repulsion in the sense
that $A/(3\pi) > B$---a condition that we have shown always obtains
for sub-thermal masses accreting at the Bondi rate, and for sufficiently low-mass super-thermal masses accreting at the Hill rate ($M < M_{\rm repulsion,visc}$)---repulsion does not much affect
the gap surface density $\Sigma_{\rm p}$. Figure \ref{fig:m_final_visc} demonstrates that different
choices for the repulsion coefficient 
$B = \{10^{-2}, 10^{-3}, 10^{-4}\}\times A_{\rm Bondi}$ 
all yield practically the same $\Sigma_{\rm p}$
(when corrected to the true sub-grid value)
relative to $\Sigma_+$. What repulsion, 
in combination with consumption, affects instead 
is how much gas leaks past the planet
into the inner disc: the three different values
for $B$ in Figure \ref{fig:m_final_visc} yield three inner disc
surface densities that, from equation (\ref{eqn:mdot-0}), scale as $\Sigma_-/\Sigma_+ \simeq (1+B/\nu)/[1 + A_{\rm Bondi}/(3\pi \nu)]$. This factor scales as $3\pi B/A_{\rm Bondi}$ when
$A_{\rm Bondi}/(3\pi) > B > \nu$ (dot-dashed and dotted lines), and as $1/[1 + A_{\rm Bondi}/(3\pi\nu)]$ when $B < \nu$ (solid line; in this limit repulsion has no effect).

\section{Viscous discs: Gas Giant Growth}\label{sec:mass}

\subsection{Numerical calculation at $r_{\rm p} = 10$ au}\label{sec:nc10au}
We now relax the assumption that the planet mass remains
fixed, and at every timestep update $M_{\rm p}$ 
according to $\dot{M}_{\rm p}$ computed
using equation (\ref{eqn:subgrid}). Our numerical procedure 
is unchanged from section \ref{sec:num}
except that we initialize the planet
mass at $M_{\rm p}(0) = 0.1 \, M_{\rm J}$
and allow it to grow. For our nominal
disc parameters ($\alpha = 10^{-3}$, $h = 0.054$ at $r_{\rm p} = 10$ au),
a starting planet mass of $0.1 M_{\rm J}$ ($m \simeq 0.95 \times 10^{-4}$)
implies that, initially, $A = A_{\rm Bondi}$, $A_{\rm Bondi}/(3\pi B) \simeq 1.3/h \simeq 24$ (a consumption-dominated
gap), $A_{\rm Bondi}/(3\pi \nu) \simeq 19$ (a strong outer
gap contrast),
and $B/\nu \simeq 0.79$ (a weak inner gap contrast).

\begin{figure}
    \centering
    \includegraphics[width=\linewidth]{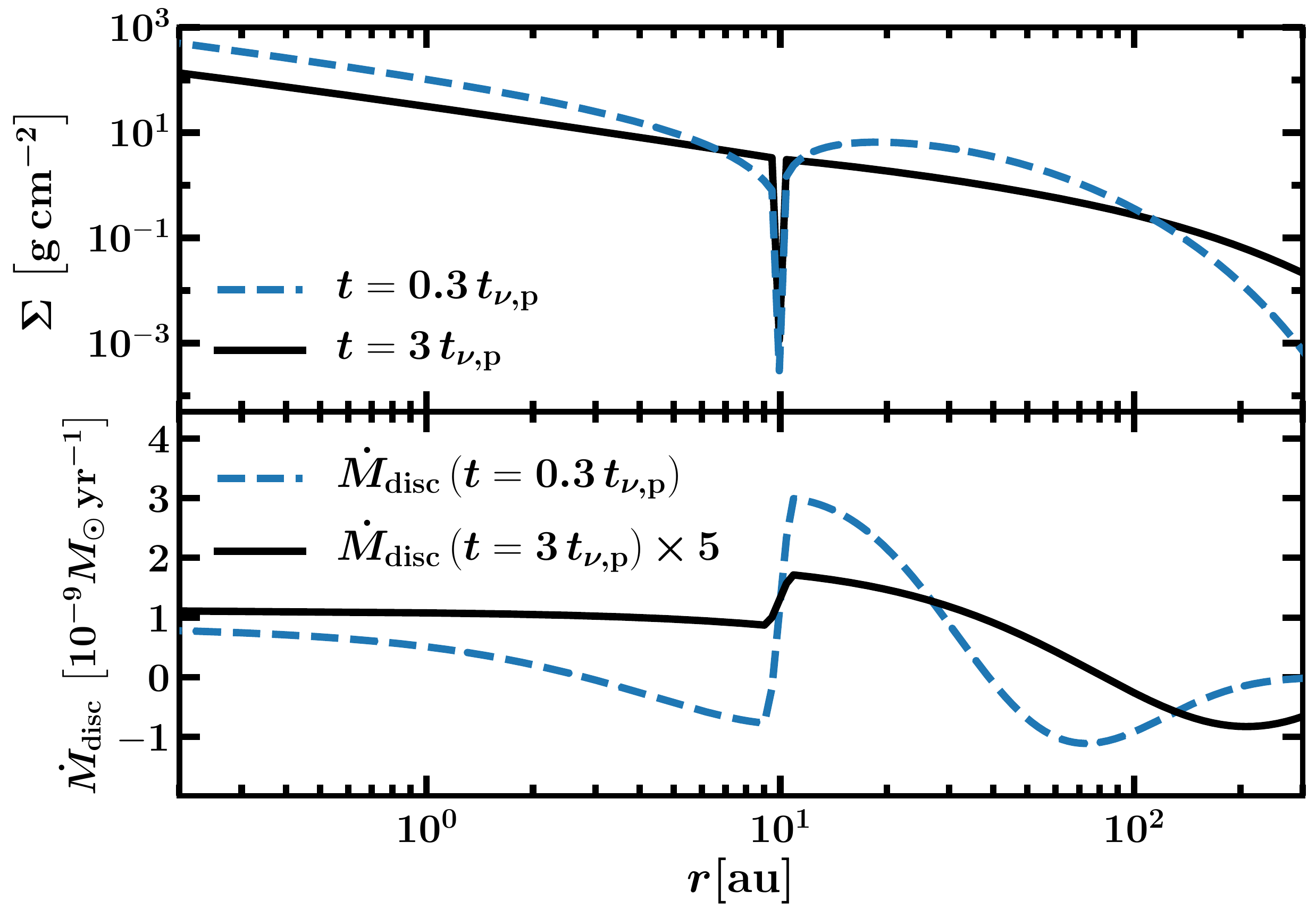}
    \caption{Snapshots of the surface density profile $\Sigma(r)$ and disc accretion rate $\dot{M}_{\rm disc}(r) = -2 \pi \Sigma u_r r$ ($> 0$ for accretion toward the star)
    for a planet embedded at $r_{\rm p} = 10$ au in a viscous $\alpha = 10^{-3}$ disc.
    The planet mass is allowed to freely grow starting from $M_{\rm p}(0) = 0.1 M_{\rm J}$. 
    At $t = 0.3 t_{\nu,{\rm p}}$, the planet resides in
    a consumption-dominated, asymmetric gap
    (top panel, dashed curve) and accretes from regions
    both exterior and interior to its orbit which have
    not yet viscously relaxed (bottom panel, dashed curve).
    At the later time $t = 3 t_{\nu,{\rm p}}$, 
    the planet has grown sufficiently (see also Figure \ref{fig:Mp_vs_t_visc}) that its gap is now repulsion-dominated
    and more symmetric (top panel, solid curve);
    the planet now
    accretes only from the outer disc, reducing the flow
    of mass into the inner disc by less than a factor of 2
    (bottom panel, solid curve. At this time
    we have multiplied $\dot{M}_{\rm disc}$ by 
    a factor of 5 for easier viewing).}
    \label{fig:sigma_mdot_M0_10_MJ}
\end{figure}

Figure \ref{fig:sigma_mdot_M0_10_MJ} shows two snapshots in time of
$\Sigma(r)$ and the disc mass flow rate
$\dot{M}_{\rm disc}(r) = -2 \pi \Sigma r u_r$, where 
\begin{align}
    u_r = -\frac{3}{\Sigma r^{1/2}} \frac{\partial}{\partial r} \left(\nu \Sigma r^{1/2} \right)
\end{align}
is the gas radial velocity (e.g., \citealt{fkr_2002})
evaluated numerically from our 
solution for $\Sigma$ (omitting the single-point discontinuity at $r = r_{\rm p}$).
Note that $\dot{M}_{\rm disc} > 0$ indicates inward
mass transport, toward the star. The planet accretes predominantly from the outer disc, notwithstanding a small contribution from the inner disc 
before the disc has viscously relaxed;
this early-time contribution can be 
seen at $t = 0.3\,t_{\nu,{\rm p}}$ 
when $\dot{M}_{\rm disc} < 0$ from 
$r \sim 3$ au to the planet's orbit.  The behaviour of $\dot{M}_{\rm disc}$ at $r \sim 100$ au is 
characteristic of a viscous disc near its turn-around ``transition
radius'' (\citealt{lp_1974}; \citealt{hcg_1998}), outside of which the disc has not yet viscously
relaxed; this outermost disc behaviour is not
caused by the planet.

Embedded in
Figure \ref{fig:sigma_mdot_M0_10_MJ}
is our assumption, first mentioned
in section \ref{sec:proc}, that the disc
flow inside the gap maintains continuity.
At $t = 3 t_{\nu,{\rm p}}$, $\dot{M}_{\rm disc} (r > r_{\rm p})$ is, to within a factor of 2, the same as $\dot{M}_{\rm disc} (r < r_{\rm p})$. 
Because the gap surface density $\Sigma_{\rm p}$ at this time is
about 4 orders of magnitude smaller than
the surface densities $\Sigma_+$ and $\Sigma_-$ outside the gap, the
radial velocity $|u_r|$ within the gap
must be 4 orders of magnitude larger
than the radial velocities outside,
to maintain the near-constancy
of $\dot{M}_{\rm disc}$ across $r_{\rm p}$.
Since the radial accretion
velocities away from
the gap are of order $r/t_\nu \sim \nu/r \sim \alpha h c_{\rm s} \sim 2$ cm/s,
we must have $|u_r| \sim 0.2$ km/s within
the gap. How such
a radial velocity is achieved is 
not specified by our model, which 
does not resolve the gap spatially.

Figure \ref{fig:Mp_vs_t_visc} displays
the planet's mass as a function of time. 
We identify
a consumption-dominated phase during which the planet
grows from 0.1 to $5 \,M_{\rm J}$ ($M < M_{\rm repulsion,visc}$;   
equation \ref{eqn:repulsionmass}) 
and a slower repulsion-dominated phase between
5 and $8 \,M_{\rm J}$ ($M > M_{\rm repulsion,visc}$). 
During the first phase, accretion starts at the Bondi
rate and switches to the Hill rate once $M_{\rm p} > M_{\rm thermal} \simeq 0.5 \,M_{\rm J}$ (equations \ref{eqn:thermalmass}
and \ref{eqn:bondi}--\ref{eqn:hill}).
A consumption-dominated ($A/(3\pi) > B$) and deep ($A/(3\pi) > \nu$) gap implies from
(\ref{eq:m_dot_ratio}) that 
$\dot{M}_{\rm p} \simeq \dot{M}_+$, i.e., the planet's accretion rate 
is about as large as it can be.
During the final repulsion-limited phase, when
$M_{\rm p} > 5 \,M_{\rm J}$ and
$A_{\rm Hill}/(3\pi) > B$,
consumption slows and the planet undergoes a last
near-doubling in mass as the remainder of the disc
diffuses away, onto the star.

\subsection{Analytic estimates of the final planet mass}
\begin{figure}
    \includegraphics[width=\linewidth]{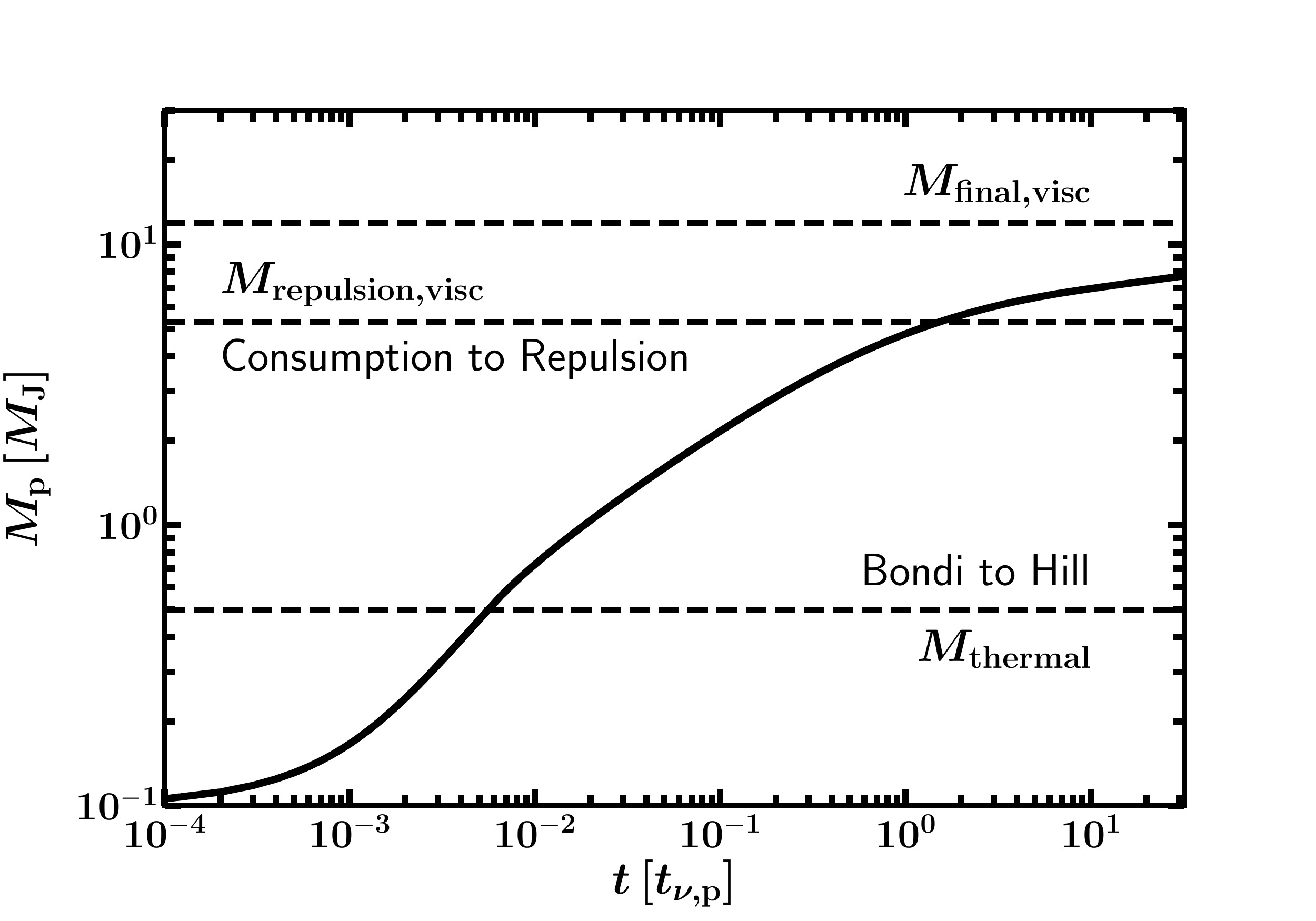}
    \hfill
    \caption{Accretion history of a planet of initial mass $M_{\rm p}(0) = 0.1 M_{\rm J}$ embedded at $r_{\rm p} = 10$ au (where $h = 0.054$) in a viscous disc of initial mass $M_{\rm disc} = 15.5 M_{\rm J}$. Transitions from Bondi
    accretion to Hill accretion ($M_{\rm thermal}$, equations \ref{eqn:bondi}--\ref{eqn:hill} and \ref{eqn:thermalmass}), and from consumption to repulsion-dominated gaps ($M_{\rm repulsion,visc}$, equation \ref{eqn:repulsionmass}), are indicated. 
    An analytic estimate
    of the final planet mass is plotted
    as $M_{\rm final,visc}$ (equation \ref{eqn:asymsolvisc}), computed assuming repulsion-dominated conditions (at $r_{\rm p} = 10$ au for this disc mass, conditions are actually intermediate between the
    repulsion and consumption limits, and so plotting equation \ref{eqn:consume1} which assumes consumption-dominated conditions would give a similar result as equation \ref{eqn:asymsolvisc}; see also Figure \ref{fig:m_final_r_visc}).}
    \label{fig:Mp_vs_t_visc}
\end{figure}

We can compare our numerical result for the final mass at $r_{\rm p} = 10$ au to the following analytic estimates, derived by 
neglecting the initial short-lived Bondi accretion phase and
assuming that at all times the planet accretes at the Hill rate ($A=A_{\rm Hill}$) 
and has a large inner gap contrast ($B >\nu$):
\begin{align}
\dot{m} &= \frac{A_{\rm Hill} \Sigma_{\rm p}}{M_\star} \nonumber \\
&= \frac{A_{\rm Hill}}{M_\star} \frac{\Sigma_+ \nu}{A_{\rm Hill}/(3\pi) + B}
\end{align}
where we have used (\ref{eqn:oom2}). At small orbital
distances, final planet masses exceed
$M_{\rm repulsion,visc}$
and so their final growth phase is repulsion-limited:
\begin{align}\label{eqn:viscmdot}
\dot{m} & = \frac{A_{\rm Hill}}{B} \frac{\Sigma_+ \nu}{M_\star} \nonumber \\
    & = 55 \alpha h^5  m^{-4/3} \frac{\Sigma_+ r_{\rm p}^2}{M_\star} \Omega \,.
\end{align}
We approximate $\Sigma_+$
using the similarity
solution for an isolated viscous disc with no planet and
$\nu \propto r^1$:
\begin{align} \label{eqn:simtime}
    \Sigma_+ \sim \frac{M_{\rm disc}}{2\pi r_1^2} \left(\frac{r_1}{r_{\rm p}}\right) T^{-3/2} e^{ -(r_{\rm p}/r_1)/T}
\end{align}
where $T \equiv 1 + t/t_1$, $t_1 \equiv r_1^2/[3 \nu(r_1)]$, and $M_{\rm disc}$ is the initial disc mass (\citealt{lp_1974}; \citealt{hcg_1998}). 
Integrating equation (\ref{eqn:viscmdot})
from $t=0$ to $t$ gives
\begin{align} \label{eqn:fullsolvisc}
    m(t) \sim & \left(\frac{385}{18\sqrt{\pi}}   \frac{M_{\rm disc}}{M_\star} \frac{h^5}{h_1^2}  \frac{r_1}{r_{\rm p}} \right)^{3/7} \nonumber \\   &\times  \left[\text{Erf}\left(\sqrt{\frac{r_{\rm p}}{r_1}}\right)-\text{Erf}\left(\sqrt{\frac{r_{\rm p} t_1}{r_1 (t + t_1)}}\right)\right]^{3/7} (\repulsionlimited)
\end{align}
where $h_1$ is the disc aspect ratio at $r_1$.
As $t \rightarrow \infty$, equation (\ref{eqn:fullsolvisc}) simplifies to
\begin{align} \label{eqn:asymsolvisc}
    m_\mathrm{final,visc} \sim& \left[\frac{385}{18\sqrt{\pi}}   \frac{M_{\rm disc}}{M_\star} \frac{h^5}{h_1^2}  \frac{r_1}{r_{\rm p}}
    \text{Erf}\left(\sqrt{\frac{r_{\rm p}}{r_1}}\right)
    \right]^{3/7} (\repulsionlimited)
    \end{align}
which further simplifies in the limit $r_{\rm p} \ll r_1$ (away from the initial disc outer edge) to
\begin{align}\label{eqn:solviscsimple}
    M_{\mathrm{final,visc}} &\sim 10 \, M_\mathrm{J}  \, \left( \frac{M_{\rm disc}}{15.5 \,M_{\rm J}} \right)^{3/7} \left( \frac{r_{\rm p}}{10 \, {\rm au}} \right)^{9/28} \,\, (\repulsionlimited)
\end{align}
for our fiducial parameters. Note that $M_{\rm final,visc}$ in these limits is independent of $\alpha$ and $r_1$. Equation (\ref{eqn:solviscsimple}) may be reproduced to order-of-magnitude by multiplying $\dot{m}$ (evaluated at $t = t_1$) by $t_1$. In Figure \ref{fig:Mp_vs_t_visc} we plot
equation (\ref{eqn:asymsolvisc})
as the uppermost horizontal
dashed line, labeled $M_{\rm final,visc}$. 

At the largest orbital distances,
conditions tend to remain consumption-limited as
$M_{\rm p}$ stays below $M_{\rm repulsion,visc}$.
Then the planet accretes
nearly all of the disc gas that tries to diffuse
past the planet---and diffusion can be in the outward
direction ($\dot{M}_{\rm disc} < 0$) if the planet is located near or beyond
the disc's turn-around radius. Accordingly
we estimate the planet mass as
\begin{align}
M_{\rm p}(t) \sim \int^t_0 |\dot{M}_{\rm disc}| dt \,\,\,\,\,\,\,\,\,\,\,\,\,\,\,\,\,\,\,\,\,\,\,\,\,\,\,\,\,\,\,\,\,\,\,\,\,\,\,\,\,\,\,\,\,\,\,\,\,\,\,\,\,\, (\consumptionlimited) 
\end{align}
where $\dot{M}_{\rm disc}$
is approximated by the no-planet similarity solution (equation 35
of \citealt{hcg_1998}). For $r_{\rm p}\leq r_1/2$,
\begin{align} \label{eqn:consume1}
M_{\rm p}(t) &\sim M_{\rm disc} \left( e^{-r_{\rm p}/r_1} - \frac{e^{-(r_{\rm p}/r_1)/T}}{\sqrt{T}} \right) \,\,\,\,\,\,\,\,\,\,\,\,\,\,\,\, (\consumptionlimited)
\end{align}
and for $r_{\rm p}>r_1/2$,
\begin{align} \label{eqn:consume2}
M_{\rm p}(t) \sim M_{\rm disc} \left( \sqrt{\frac{2r_1}{r_{\rm p}}} e^{-1/2} - e^{-r_{\rm p}/r_1} - \frac{e^{-(r_{\rm p}/r_1)/T}}{\sqrt{T}} \right) \nonumber \\
(\consumptionlimited).
\end{align}
We will make use of equations (\ref{eqn:fullsolvisc}), (\ref{eqn:consume1}), and (\ref{eqn:consume2})
in section \ref{sec:sum} when we discuss, in the context
of observations, how the final planet
mass depends on disc mass and orbital distance.

\newpage
\section{Planets in Inviscid Wind-Driven Discs}\label{sec:windy}
Motivated by recent ALMA observations that point to
little or no turbulence in protoplanetary discs
(e.g., \citealt{p_etal_2016}; \citealt{f_etal_2017}),
and by theoretical work arguing that discs are,
for the most part, laminar because they are too cold and dusty
to support magnetorotational turbulence (e.g., \citealt{g_1996};
\citealt{pc_2011}; \citealt{b_2011}), we here turn away from the $\alpha$-based picture of turbulent
and diffusive discs, and consider instead inviscid (zero
viscosity) discs that accrete by virtue of magnetized winds (e.g.,
\citealt{b_etal_2016}; \citealt{b_2016}). 
We review how wind-driven discs
work in section \ref{sec:wind} and
how planets open repulsive gaps in inviscid discs
in section \ref{sec:inv}. 
We then study how repulsion combines with consumption
to set gap depths and planetary accretion
rates, analytically in section \ref{sec:invoom}
and numerically in section \ref{sec:numwind}.

\subsection{Wind-driven accretion discs}\label{sec:wind}
Inviscid, wind-driven accretion discs do not behave
diffusively. Instead they are governed by simple advection:
at every radius $r$, material moves inward
with a vertically-averaged radial speed $u_r$ because
it has lost angular momentum to a magnetized wind. 
The mass carried away by the wind itself is small
compared to the mass advected inward through the disc (see
Appendix \ref{sec:winds}, in particular
the discussion below equation \ref{eqn:lambda}).
Then from continuity,
including our planetary mass sink,
\begin{equation} \label{eqn:cont2}
\frac{\partial \Sigma}{\partial t} = \frac{1}{ r}  \frac{\partial}{\partial r} \left( \Sigma r u_r \right) - \frac{\dot{M}_{\rm p}(t)}{2\pi r} \delta (r - r_{\rm p}) \,.
\end{equation}
In Appendix \ref{sec:winds} we show how a wind-driven disc
inspired by \citet{b_etal_2016} and \citet{b_2016}
can have $u_r$ approximately constant ($<0$ for accretion).
We utilize here, for simplicity,
a constant $u_r \equiv c < 0$ model:
\begin{equation} \label{eqn:cont3}
\frac{\partial \Sigma}{\partial t} = \frac{c}{r} \frac{\partial}{\partial r} (\Sigma r) - \frac{\dot{M}_{\rm p}(t)}{2\pi r} \delta (r - r_{\rm p}) \,.
\end{equation}

It is instructive to examine the solution to (\ref{eqn:cont3})
when $\dot{M}_{\rm p}=0$. The no-planet solution is separable: 
\begin{equation} \label{eqn:sol}
\Sigma(r,t) = f(r) g(t) = \frac{M_{\rm disc}}{2\pi (ct_{\rm adv})^2} \frac{|c|t_{\rm adv}}{r} e^{-r/(|c|t_{\rm adv})} e^{-t/t_{\rm adv}}
\end{equation}
for constants $M_{\rm disc}$ (the initial disc mass) and $t_{\rm adv}$, which we interpret as a disc radial advection time or drain-out time.
For $c = -4$ cm/s (a value we relate
to magnetic field parameters in Appendix \ref{sec:winds})
and $t_{\rm adv} = 3$ Myr,
the characteristic disc size is
$|c|t_{\rm adv} \simeq 25$ AU, which seems reasonable.
Equation (\ref{eqn:sol}) resembles the \citet{lp_1974}
 solution for a viscous disc which gives, for $\nu \propto r^1$, 
a surface density profile that scales as $r^{-1} \exp(-r/r_1)$ at
fixed $t$ (equation \ref{eqn:simtime}). 
This spatial resemblance is not
surprising, as our viscous
disc happens also to have an 
accretion velocity that is constant
with radius: $|u_r| \sim r/t_\nu \sim \nu/r =$ constant.
However,
the solutions differ in their time behaviours;
at fixed $r$,
the wind-driven surface density decays exponentially as $\exp(-t/t_{\rm adv})$, whereas our viscous disc decays
as a power law $t^{-3/2}$ (within viscously relaxed
regions at small radii; \citealt{lp_1974, hcg_1998}).
Viscous discs evolve
more slowly because they conserve their total angular
momentum; they can only drain away on the inside by 
redistributing their angular momentum to the outside
in a kind of zero-sum game. Wind-driven discs are
not so constrained; they lose their angular momentum
wholesale to a wind, and so can dissipate more quickly.

We emphasize that $u_r = c$
is a vertically averaged, mass weighted, 
radial accretion velocity. In simulations by
\citet{bs_2013} of discs whose magneto-thermal
winds are anchored at their electrically conductive
surfaces, accretion actually occurs 
in a vertically thin, rarefied layer 
several scale heights above the midplane.
The radial accretion velocity in this high-altitude layer is fast,
on the order of the sound speed
$c_{\rm s}$. The bulk of the mass
of the disc, below this layer, is inert
(see fig.~10 of \citealt{bs_2013}).
It is with this static and inviscid gas, extending from the midplane to a couple scale heights above and below, that the planet
interacts, as we now describe.

\subsection{Repulsion in inviscid discs}\label{sec:inv}
Without viscosity, disc gas in the vicinity
of the planet depletes indefinitely, as it
is repelled by the planetary Lindblad
torque but cannot diffuse back.
Under these conditions, \citet{gc_2019a} derived
how the gas density at the center of
the planet's gap scales with elapsed
time $t$, for a given planet-to-star mass
ratio $m = M_{\rm p}/M_\star$ and disc
aspect ratio $h = H/r$ (see the inviscid branch
of their equation 17, and also their appendix):
\begin{align} \label{eqn:inv}
\frac{\Sigma_{\rm p}}{\Sigma_-} \sim h^{549/49} m^{-4} (\Omega t)^{-39/49} \equiv \widetilde{B}_{\rm inv}^{-1}
\end{align}
where $\Omega$ is the orbital frequency of the planet, $\Sigma_{\rm p}$ is the surface density within the gap, and $\Sigma_-$ is the surface density
downstream of
the planet in the accretion flow (see Figure \ref{fig:sig_cartoon}).
By construction, $t$ is the time over which the planet's mass is close to its given value $m$ (say within a factor of 2).
In practice, for inviscid discs where gaps deepen 
dramatically 
with increasing planet mass, the mass doubling time of a planet lengthens with each doubling, so $t$ is of order the system age.

Equation (\ref{eqn:inv}) does not apply when
$\widetilde{B}_{\rm inv} < 1$, i.e., when a repulsive gap has not yet been opened because not enough time has elapsed for
a given planet mass. 
To account for this possibility, we generalize
(\ref{eqn:inv}) using
\begin{equation}\label{eqn:invgen}
\frac{\Sigma_{\rm p}}{\Sigma_-} \sim \frac{1}{1+\widetilde{B}_{\rm inv}}
\end{equation}
by analogy with equation (\ref{eqn:oom1})
for the viscous case.
Note that $\widetilde{B}_{\rm inv}$ is dimensionless while its viscous counterpart $B$ has dimensions of viscosity.

\subsection{Consumption and repulsion combined}\label{sec:invoom}
We now assemble the physical ingredients laid out in
sections \ref{sec:wind} and \ref{sec:inv}
into a sketch of how
consumption and repulsion combine in an inviscid,
wind-driven disc. Following by analogy 
our analysis in section \ref{sec:oom} for a
viscous disc, 
we first write down mass conservation (see
equation \ref{eqn:mass} and Figure \ref{fig:sig_cartoon}):
\begin{align} \label{eqn:massinv}
\dot{M}_+ &= \dot{M}_- + \dot{M}_{\rm p} \nonumber \\
2\pi \Sigma_+ r |c| &= 2\pi \Sigma_- r |c| + \dot{M}_{\rm p} \nonumber \\
&= 2\pi \Sigma_- r |c| + A \Sigma_{\rm p}  
\end{align}
where in lieu of the viscosity we now have
$r |c|$. After replacing $\Sigma_-$ in (\ref{eqn:massinv})
using our momentum relation (\ref{eqn:invgen}), we have
\begin{align}
\Sigma_+  \sim (1+\widetilde{B}_{\rm inv})\Sigma_{\rm p} + \frac{A}{2\pi r|c|} \Sigma_{\rm p}
\end{align}
which implies the outer gap contrast
\begin{align} \label{eqn:outer2}
\frac{\Sigma_{\rm p}}{\Sigma_+} \sim \frac{1}{1+A/(2\pi r|c|) + \widetilde{B}_{\rm inv}} \,.
\end{align}
As in the viscous case (equation \ref{eqn:oom2}),
we see here that consumption ($A/(2\pi r|c|)$)
and repulsion ($\widetilde{B}_{\rm inv}$) add. Taking
$A$ to be the Bondi value (equation \ref{eqn:bondi})
gives the ratio
\begin{align}\label{eqn:sensitive}
\frac{A_{\rm Bondi}/(2\pi r|c|)}{\widetilde{B}_{\rm inv}} \sim &  \,\frac{0.5}{2\pi} \frac{h^{353/49}}{m^{2} (\Omega t)^{39/49}} \frac{\Omega r_{\rm p}}{|c|} \nonumber \\
\sim & \,0.04 \left( \frac{M_{\rm p}}{0.1 \, M_{\rm J}} \right)^{-2} \left( \frac{t}{3 \, {\rm Myr}}  \right)^{-39/49} \times \nonumber \\
 & 
 \left( \frac{|c|}{4 \, {\rm cm/s}} \right)^{-1} \left( \frac{r_{\rm p}}{10 \, {\rm au}} \right)^{489/196}
\end{align}
which informs us that  
repulsion dominates consumption ($\widetilde{B}_{\rm inv} > A_{\rm Bondi}/(2\pi r|c|)$) when
\begin{align}
M_{\rm p} > M_{\rm repulsion,inv} \sim & \, 0.02 \,M_{\rm J} \left( \frac{t}{3 \, {\rm Myr}}  \right)^{-39/98} \times \nonumber \\
 & 
 \left( \frac{|c|}{4 \, {\rm cm/s}} \right)^{-1/2} \left( \frac{r_{\rm p}}{10 \, {\rm au}} \right)^{489/392} \,.
\end{align}
That repulsion dominates consumption
even for small masses is in contrast 
to the viscous case (see equation \ref{eqn:repulsionmass} for
$M_{\rm repulsion,visc}$). Repulsion-dominated gaps are 
symmetric between the inner and outer discs
(equations \ref{eqn:invgen} and \ref{eqn:outer2}):
\begin{align}
\Sigma_{\rm p}/\Sigma_-  & \sim   \Sigma_{\rm p}/\Sigma_+ \sim  1/(1+\widetilde{B}_{\rm inv})  \nonumber \\
& \sim  2 \times 10^{-3} \left( \frac{h}{0.054} \right)^{549/49} \left( \frac{10^{-4}}{m} \right)^4 \left( \frac{3 \, {\rm Myr}}{t} \right)^{39/49}  \label{eqn:symmetric}
\end{align}
where for the last equality we have assumed
that the gaps are deep ($\widetilde{B}_{\rm inv} > 1$).
Under these conditions,
we may estimate a final accreted planet mass
by time-integrating
\begin{align}
\dot{M}_{\rm p} &= A_{\rm Bondi} \Sigma_{\rm p} \nonumber \\
&\sim A_{\rm Bondi} \frac{\Sigma_+}{\widetilde{B}_{\rm inv}} \nonumber \\
& \sim \frac{A_{\rm Bondi}}{\widetilde{B}_{\rm inv}} \frac{M_{\rm disc}}{2\pi (ct_{\rm adv})^2} \frac{|c|t_{\rm adv}}{r_{\rm p}} e^{-r_{\rm p}/(|c|t_{\rm adv})} e^{-t/t_{\rm adv}} \label{eqn:int}
\end{align}
from $t=0$ to $\infty$, 
where for $\Sigma_+$ we have employed the no-planet solution (\ref{eqn:sol}). This last approximation is analogous
to the one we made in (\ref{eqn:simtime}) for a viscous disc. Equation (\ref{eqn:int}) integrates to yield
\begin{align} \label{eqn:m_final_wind_full}
M_\mathrm{final,inv} & \sim  \left[\frac{1.5}{2\pi} \, \Gamma\left(\frac{10}{49}\right) \left(\frac{M_{\rm disc}}{M_\star} \right) \left(\frac{r_{\rm p}}{ |c|t_{\rm adv}} \right)  \right. \nonumber \\ &  \times \left. h^{353/49}(\Omega t_{\rm adv})^{10/49}   e^{-r_{\rm p}/(|c|t_{\rm adv})} \right]^{1/3} M_\star \nonumber \\
& \sim  0.3 \, M_{\rm J} \left( \frac{M_{\rm disc}}{15.5 \,M_{\rm J}} \right)^{1/3} \left( \frac{r_{\rm p}}{10 \, {\rm au}} \right)^{163/196} e^{-r_{\rm p}/(3|c|t_{\rm adv})}\nonumber \\
& (\repulsionlimited)
\end{align}
where $\Gamma$ is the gamma function, and
the numerical evaluation uses our fiducial parameters including 
$|c| = 4 $ cm/s, $t_{\rm adv} = 3$ Myr, and $M_\star = 1 M_\odot$. 
Our estimated final mass of $0.3 \, M_{\rm J}$ 
at $r_{\rm p} = 10$ au remains smaller than
$M_{\rm thermal} \simeq 0.5 M_{\rm J}$ and so our use of $A_{\rm Bondi}$
is self-consistent.

Our expression (\ref{eqn:m_final_wind_full}) for $M_{\rm final,inv}$ 
resembles equation (19) of \citet{gc_2019a};
ours is an improvement as we have accounted explicitly 
for the transport properties of the disc
through the radial velocity $c$ (see the discussion of transport-limited
accretion in their section 4.1).

\subsection{Numerical simulations}\label{sec:numwind}

We test the ideas in section \ref{sec:invoom}
by numerically solving the continuity
equation (\ref{eqn:cont3})
and the momentum equation (\ref{eqn:invgen}).
To model the planetary mass sink in equation
(\ref{eqn:cont3}), we utilize the same sub-grid procedure
of section \ref{sec:num}, replacing
equation (\ref{eqn:subgrid}) with
\begin{align} \label{eqn:m_dot_win}
\dot{M}_{\rm p}(t) = A \times \frac{\Sigma(r_{\rm p},t)}{1+ \widetilde{B}_{\rm inv}} \,\,\,\, \,\,\,\, \,\,\,\, \,\,\,\, \,\,\,\, \,\,\,\, \,\,\,\, \,\,\,\, \,\,\,\, \,\,\,\, \,\,\,\, \,\,\,\, \,\,\,\, \,\,\,\, 
({\rm simulation})
\end{align}
where $\Sigma(r_{\rm p},t)$ is the grid-level
surface density in the bin containing the planet,
and $A$ and $\widetilde{B}_{\rm inv}$ are given by
equations (\ref{eqn:bondi})--(\ref{eqn:hill}) and (\ref{eqn:inv}),
respectively. The initial
mass of the planet is set to $M_{\rm p}(0) = 0.1 \, M_{\rm J}$
(we will see that using smaller initial masses hardly changes
the outcome). 
We solve the advective portion
of equation (\ref{eqn:cont3}) with a first-order
upwind scheme (e.g., \citealt{ptv_2007})
applied to a grid
that extends from $r_{\rm in} = 0.01$ au
to $r_{\rm out} = 500$ au
across 300 cells uniformly spaced in
$\log r$. We fix $c = -4$ cm/s 
and initialize the grid using (\ref{eqn:sol}),
with $t_{\rm adv} = 3$ Myr and 
$M_{\rm disc} = 15.5 M_{\rm J} = 0.015 M_\odot$, 
the same value chosen
for our viscous disc calculations. 
Our timestep is set to $\Delta t = 0.2 \Delta r_{\mathrm{min}}/|c|$, where $\Delta r_\mathrm{min} = 3 \times 10^{-3} \, \mathrm{au}$ is our smallest bin width. 
Other disc properties such as $h(r)$
and $\Omega(r)$ are the same as before.
For the outer boundary condition
we impose a ghost cell just outside
$r_{\rm out}$ where the surface density
is fixed at 0.

\begin{figure} 
    \centering
    \includegraphics[width=\linewidth]{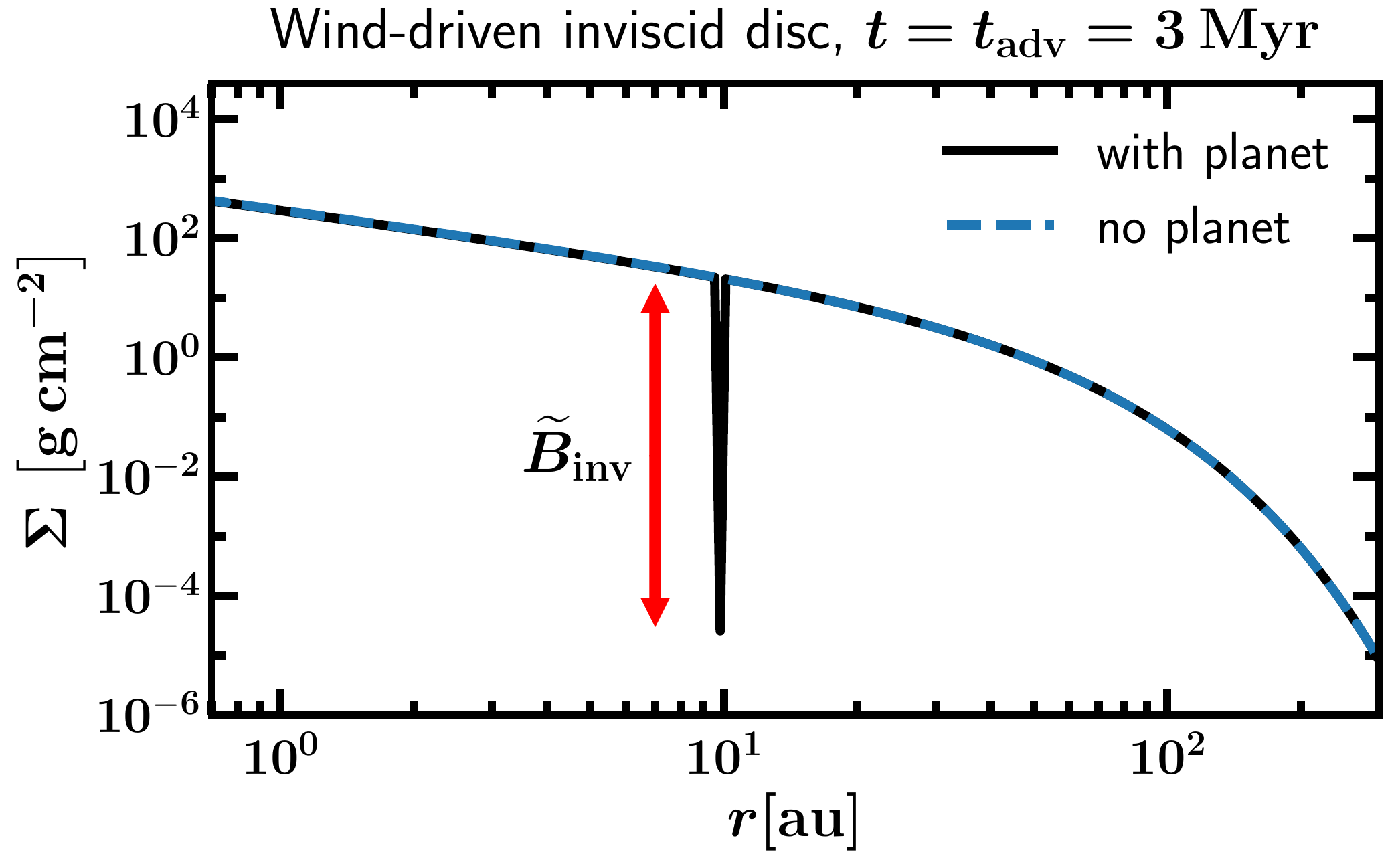}
    \caption{How the surface density profile of an inviscid disc responds to a planet that consumes disc gas
    and repels gas away by Lindblad torques. The planet,
    located at $r_{\rm p} = 10$ au, freely accretes
    starting from
    a seed mass of $0.1 M_{\rm J}$; the $\Sigma$ profile
    shown here is taken at a time $t = t_{\rm adv} = 3$ Myr, when 
    the planet has grown to $\sim$$0.3 M_{\rm J}$ (see also
    Figure \ref{fig:sigma_mdot_M0_wind}). As is the case throughout this paper,
    the planet's gap is not spatially resolved, but is modeled
    as a single cell. The ``true''
    surface density inside this cell equals the grid-level
    $\Sigma$ lowered by a factor of $\widetilde{B}_{\rm inv}$, whose
    magnitude is given by the red double-tipped arrow. The gap
    is repulsion and not consumption dominated
    ($\widetilde{B}_{\rm inv} > A/(2\pi r|c|)$, equation \ref{eqn:sensitive}); as such,
    the gap is symmetric in the sense that
    the surface density contrast with the outer disc is practically
    the same as with the inner disc. This figure is the inviscid counterpart to Figure \ref{fig:sigma_detailed} which was made for a viscous disc.}
    \label{fig:m_final_wind_tdisc_a}
\end{figure}

\begin{figure}  
    \centering
    \includegraphics[width=\linewidth]{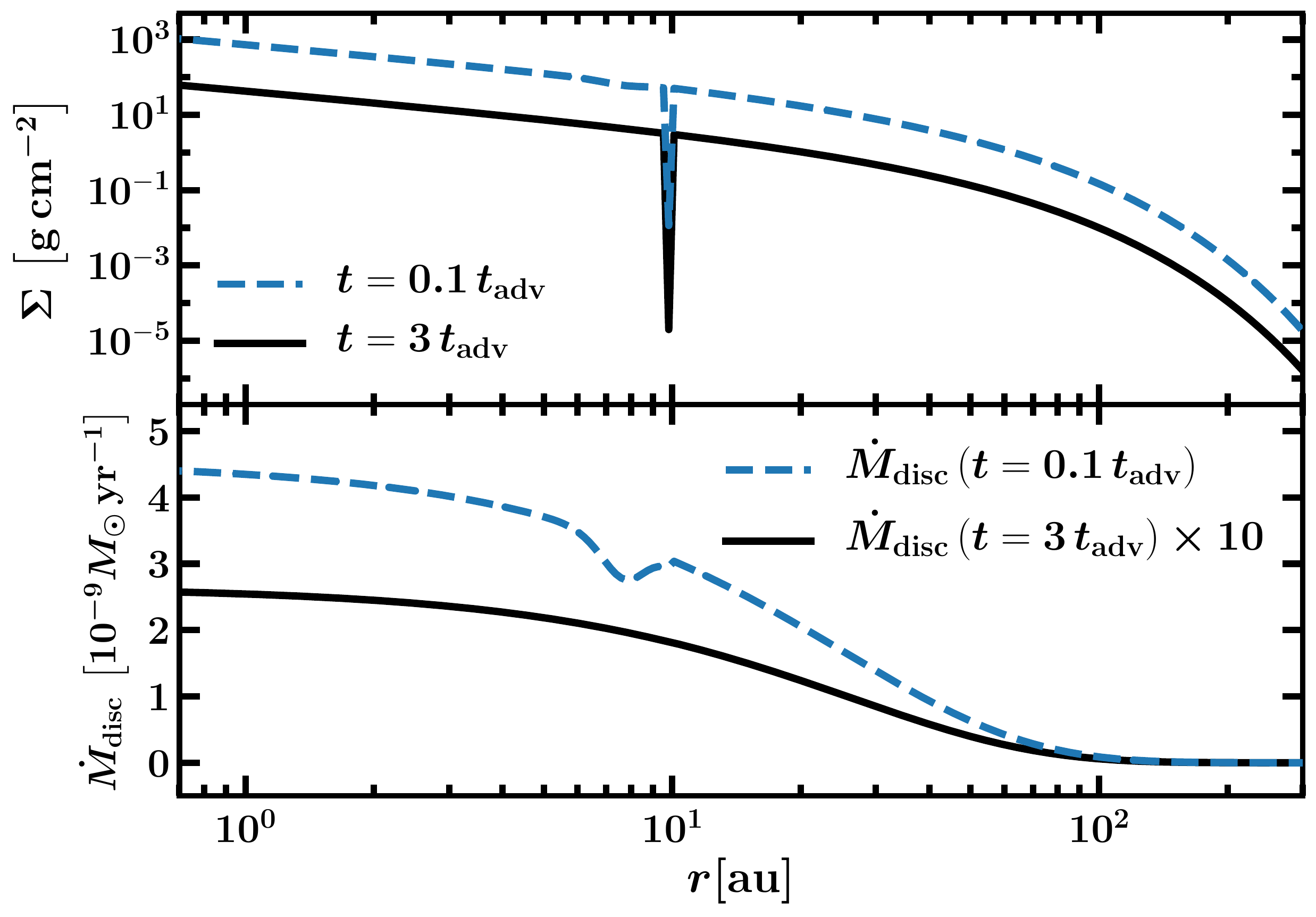}
    \caption{Snapshots of the surface density profile $\Sigma(r)$ and disc accretion rate $\dot{M}_{\rm disc}(r) = -2 \pi \Sigma u_r r$ ($> 0$ for accretion toward the star)
    for a planet embedded in an inviscid, wind-driven disc. The planet mass is allowed to freely grow starting 
    from $M_{\rm p}(0) = 0.1 \,M_{\rm J}$; the masses corresponding to the plotted times are $0.27\, M_{\rm J}$ ($t = 0.1\, t_{\rm adv} = 0.3$ Myr)
    and $0.34 \,M_{\rm J}$ ($t = 3 \,t_{\rm adv} = 9$ Myr; see also Figure \ref{fig:sigma_mdot_M0_wind}). At
    $t=3\,t_{\rm adv}$, the disc has relaxed into a quasi-steady state in
    the presence of the planetary mass sink, and $\dot{M}_{\rm disc}(r)$ looks essentially
    the same as it would without the planet; the accretion rate onto the planet is negligible compared to the disc accretion rate---the gap is repulsion-dominated---and so the disc is not materially affected.
    Even at $t = 0.1\,t_{\rm adv}$, the 
    interior surface density 
    $\Sigma_-$ and $\dot{M}_{\rm disc}$ depress
    by only $\sim$15\% because of consumption.}
    \label{fig:m_final_wind_tdisc_b}
\end{figure}

\begin{figure}
    \centering
    \includegraphics[width=\linewidth]{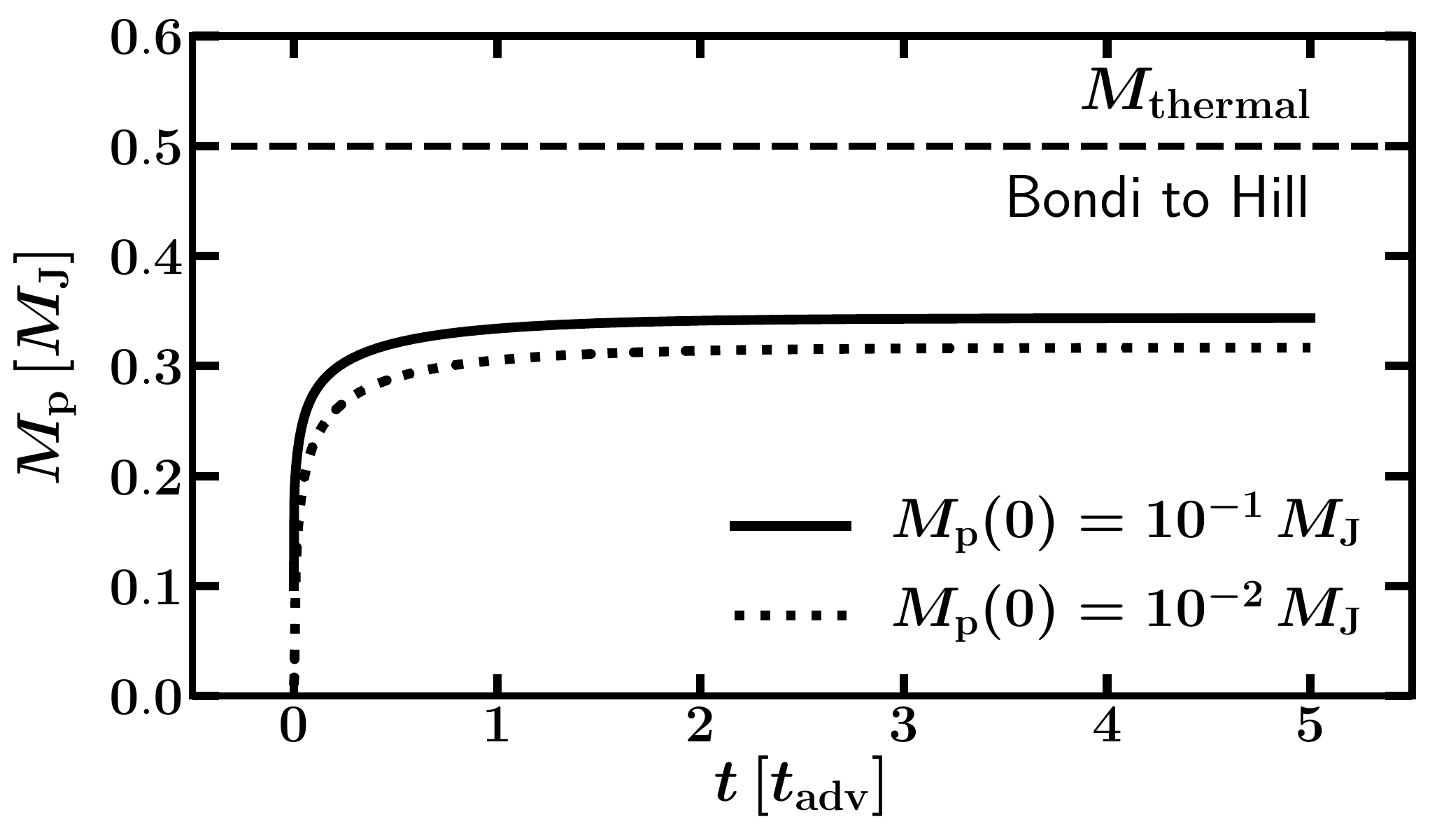}
    \caption{Mass evolution of a planet embedded at $r_{\rm p} = 10$ au in an inviscid but still accreting disc of initial mass $M_{\rm disc} = 15.5 \,M_{\rm J}$. Within $\sim$1 disc advection time $t_{\rm adv}$, the planet, whose gap is repulsion-dominated 
    ($\widetilde{B}_{\rm inv} > A_{\rm Bondi}/(2\pi r_{\rm p}|c|)$), grows to a mass of $\sim$$0.35 \, M_{\rm J}$. 
    The final planet mass varies by only $\sim$10\% when the initial seed mass $M_{\rm p}(0)$ varies by a factor of 10.
    This figure is the inviscid counterpart to Figure \ref{fig:Mp_vs_t_visc} which was made for a viscous disc.}
    \label{fig:sigma_mdot_M0_wind}
\end{figure}

Figure \ref{fig:m_final_wind_tdisc_a} (the inviscid counterpart
to Figure \ref{fig:sigma_detailed}) shows $\Sigma(r)$ at $t = t_{\rm adv}$
when $M_{\rm p}$ has grown to $0.3 \, M_{\rm J}$,
illustrating many of the features anticipated from our analytic 
treatment. Without a planet,
the surface density profile follows $r^{-1} \exp[-r/(|c|t_{\rm adv})]$ 
as expected from equation (\ref{eqn:sol}).
With a planet,
a gap is created that is nearly symmetric
between the inner and outer discs, and whose
depth is dominated by Lindblad repulsion
(enforced by our sub-grid scheme), not consumption
(equation \ref{eqn:symmetric}).
The inviscid gap is deep (scaling as $m^{-4}$;
\citealt{gs_2018}; \citealt{gc_2019a};
see also \citealt{d_2020}).
Figure \ref{fig:m_final_wind_tdisc_b} (analogous to Figure \ref{fig:sigma_mdot_M0_10_MJ})
provides snapshots of $\Sigma(r)$ and $\dot{M}_{\rm disc}(r)$ taken at different times,
and Figure \ref{fig:sigma_mdot_M0_wind} (analogous to Figure \ref{fig:Mp_vs_t_visc}) plots $M_{\rm p}(t)$.
Unlike in a viscous disc,
our example planet in an inviscid disc
does not consume most of the disc mass exterior to its orbit;
the disc accretion rate profile
$\dot{M}_{\rm disc}(r)$ is not much affected by the planet 
except during an initial transient phase at $t < t_{\rm adv}$. 
We see a need for a high radial accretion
velocity $|u_r|$ within the gap (see also section \ref{sec:nc10au}):
to ensure that $\dot{M}_{\rm disc}$ grades smoothly across the gap as shown 
in Figure \ref{fig:m_final_wind_tdisc_b}, $|u_r|$ must increase in proportion to the
gap contrast $\Sigma/\Sigma_{\rm p}$.
Inviscid gap contrasts are 
on the order of $10^5$, and so 
$|u_r| \sim 10^5 |c| \sim 4$ km/s,
comparable to the orbital velocity.
Note that simulations of planets in inviscid
discs have not reproduced the deep gaps
expected from our analytics, finding 
gap contrasts only up to a factor of $\sim$10 
(e.g., \citealt{fc_2017,mnp_2019,mnp_2020}).
On the one hand the simulations are of limited
duration and so their gaps may not have fully developed;
on the other hand, the simulations allow for 
orbital migration and hydrodynamical instabilities,
effects which may prevent gaps from becoming
too deep in reality.

That the disc accretion flow proceeds largely unimpeded
from outside to inside the planet's orbit is a consequence
of the gap being repulsion-dominated (equation \ref{eqn:mdot-0}, with
$\nu$ replaced by $r|c|$). 
The planet diverts such a small fraction of the disc
flow that it grows from 
$0.1 \, M_{\rm J}$ to only $0.3 \, M_{\rm J}$; most of the
original $15.5 \, M_{\rm J}$ contained in the disc 
drains onto the star.
Figure \ref{fig:sigma_mdot_M0_wind}
also shows that reducing 
the initial seed mass to $M_{\rm p}(0) = 0.01 M_{\rm J}$
hardly affect the final mass.

\section{Summary and Discussion}\label{sec:sum}
Planets open gaps in circumstellar discs
in two ways: by repelling material away via Lindblad torques,
and by consuming local disc gas.
Measured relative to the disc outside the planet's orbit,
the two effects are additive:
both repulsion and consumption add to
deepen the planet's gap relative to the outer disc
(see equation \ref{eqn:oom2} or \ref{eqn:outer2}). 
Relative to the inner disc, downstream of the
mass sink presented by the planet,
the gap surface density contrast 
is set by repulsion only (see equation
\ref{eqn:oom1} or \ref{eqn:invgen}).

Many planet formation studies (e.g., \citealt{tt_2016}; 
\citealt{lee_2019})
take the planet's hydrodynamically-limited 
accretion rate 
$\dot{M}_{\rm p} = \min (\dot{M}_{\rm hydro}, \dot{M}_{\rm disc})$, 
where $\dot{M}_{\rm hydro}$ is the planetary accretion 
rate computed according to the hydrodynamics of flows
in the immediate vicinity of the planet, 
and $\dot{M}_{\rm disc}$ is
the local disc accretion rate  
(the mass crossing the planet's orbital radius, per time).
Prescribing
the planet's accretion rate in this way
is equivalent to comparing consumption, as 
measured by the 
``consumption coefficient''
$A \equiv \dot{M}_{\rm p}/\Sigma_{\rm p}$, where $\Sigma_{\rm p}$ is the 
surface density inside the gap, and repulsion, as measured by
the ``repulsion coefficient'' $B \equiv T/(\Sigma_{\rm p} \Omega r^2)$, where
$T$ is the repulsive planetary torque and $\Omega r^2$ is the angular
momentum per unit mass 
(see also \citealt{tt_2016} and \citealt{tmt_2020} who use the same framework).
Under consumption-limited conditions
($A/(3\pi) > B$),
the planet's accretion rate saturates to nearly 
the disc's accretion rate:
$\dot{M}_{\rm p} = \min (\dot{M}_{\rm hydro}, \dot{M}_{\rm disc}) = \dot{M}_{\rm disc}$. Otherwise, under repulsion-limited conditions 
($A/(3\pi) < B$),
$\dot{M}_{\rm p} = \min (\dot{M}_{\rm hydro}, \dot{M}_{\rm disc}) = \dot{M}_{\rm hydro}$. 

\subsection{Final planet masses}\label{sec:final}
\begin{figure}  
    \centering 
    \includegraphics[width=\linewidth]{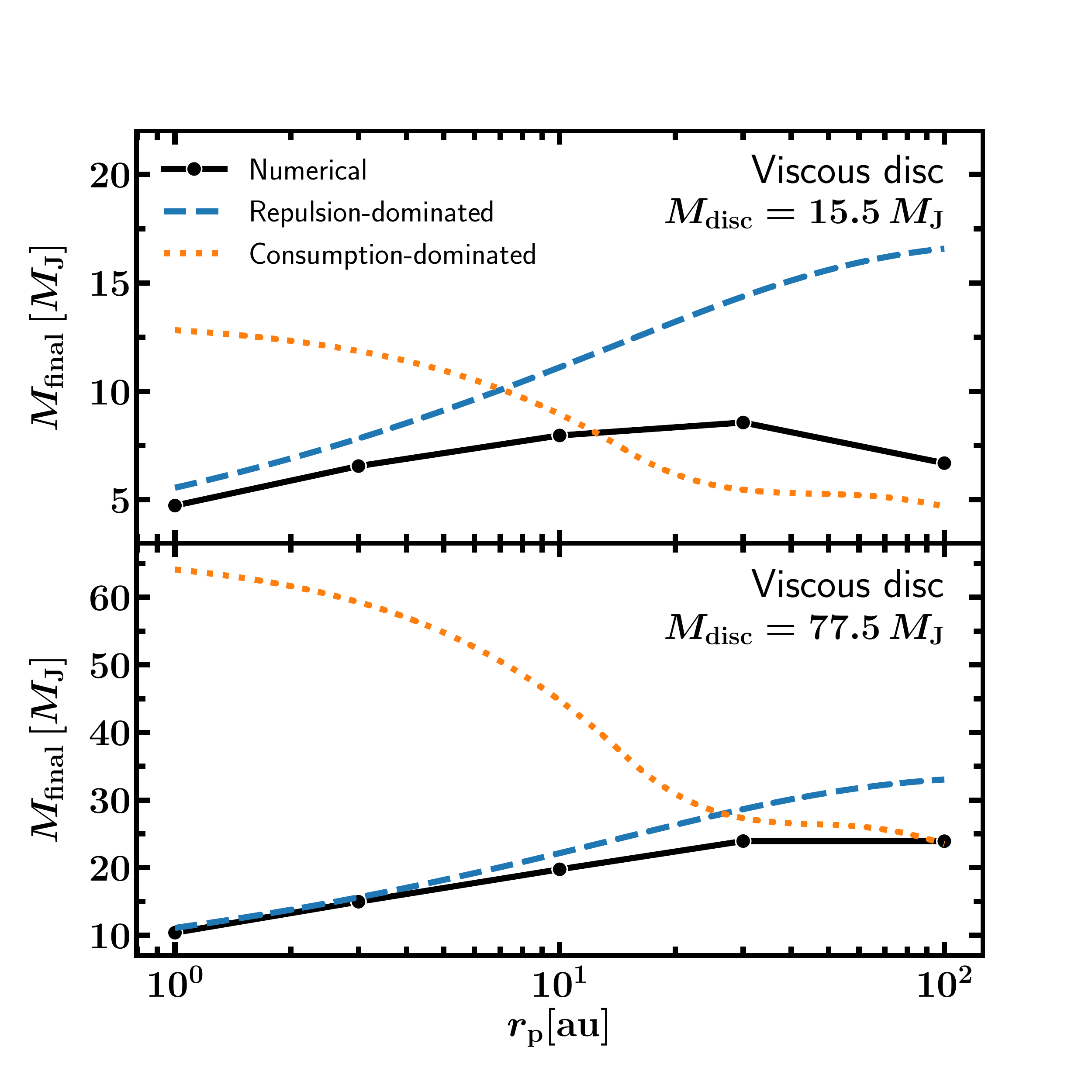}
    \caption{Final planet masses grown from viscous discs having $\alpha =
10^{-3}$ and varying total mass (top vs.~bottom panels). Planet masses are
initialized at $0.1 \,M_{\rm J}$ and grown using the 1D numerical code
of section \ref{sec:mass}, which utilizes the repulsive gap contrast of Kanagawa et al.~(\citeyear{kmt_2015};
see also \citealt{dm_2013} and \citealt{fsc_2014}) 
and gas accretion that switches from Bondi to Hill at the thermal mass. Points 
are plotted at $t = 50\,t_1 = 85$ Myr,
where $t_{\rm 1} = r_1^2/[3\nu(r_1)]$ is the viscous diffusion time at $r_1 = 30$ au. Analytic curves are
given by equation (\ref{eqn:fullsolvisc}) for the repulsion limit (dashed blue), and equations (\ref{eqn:consume1})--(\ref{eqn:consume2}) for the
consumption limit (dotted orange), also evaluated at $t = 50 t_1$.
At most orbital distances, planet mass growth
is limited by repulsion-dominated gaps; only
at the largest distances, where the disc aspect
ratio is large, are gaps relatively harder to open and conditions remain consumption-limited. The analytics, which are derived assuming the planet mass is small compared to the disc mass, are a better guide for the more massive disc in the bottom panel.}
    \label{fig:m_final_r_visc}
\end{figure}

In conventional viscous discs
with large enough $\alpha$-diffusivities\footnote{If the
Shakura-Sunyaev $\alpha \lesssim 10^{-4}$,
discs respond to planetary torques as if they were inviscid
(\citealt{gc_2019a}, their fig.~1).}
and our assumed parameters,
planets begin their growth under consumption-dominated
conditions and possibly continue their growth under repulsion-dominated
conditions, arriving at final masses well in excess of a Jupiter.
We show in Figure \ref{fig:m_final_r_visc} 
the final mass of a planet 
embedded in an $\alpha = 10^{-3}$ disc,
as a function of the planet's orbital distance $r_{\rm p}$,
computed using our numerical code of
sections \ref{sec:consumption}--\ref{sec:mass}. 
Final planet masses increase gradually from
$4 \, M_{\rm J}$ at 1 au, to $8\,M_{\rm J}$ at 30 au,  
in a disc of initial mass $M_{\rm disc} = 15.5 M_{\rm J}= 0.015 M_\odot $.
In a disc $5\times$ more massive, the corresponding range of planet masses
is 9--$20 \, M_{\rm J}$. 
The final masses are not sensitive to $\alpha$ insofar as $\alpha$ controls only the timescale over which the disc evolves (modulo disc dispersal by some other 
means, e.g., photoevaporation; 
see \citealt{tmt_2020}). 
Final masses do 
depend on the initial mass of the disc, 
scaling as $M_{\rm disc}^{3/7}$ under
repulsion-dominated conditions (equation \ref{eqn:asymsolvisc}) and $M_{\rm disc}^1$
under consumption-dominated conditions (equation \ref{eqn:consume1} or \ref{eqn:consume2}).
The trend of final planet mass with distance shown
in Figure \ref{fig:m_final_r_visc} 
follows, for the most part, the trend predicted
for repulsion-limited conditions, 
except at large $r_{\rm p}$ where 
consumption dominates. 
The final mass profiles in Figure \ref{fig:m_final_r_visc}
recall those of the super-Jupiters in the
HR 8799 system; the four planets, located between
15 and 70 AU of their host star, have practically
the same mass, about 6--7 $M_{\rm J}$ (\citealt{wgd_2018}).

Initially and everywhere in a viscous disc, a planet, despite opening a gap, consumes 
practically all of the disc gas that tries
to diffuse past its orbit (equation \ref{eq:m_dot_ratio}
with $A/(3\pi) > B > \nu$, where $\nu$ is the disc viscosity).
This consumption-limited behaviour 
persists up to a repulsion mass
$M_{\rm repulsion,visc}
\simeq 5 \, M_{\rm J} \,[r_{\rm p}/(10 \, {\rm au})]^{9/16}$
(equation \ref{eqn:repulsionmass}), above which repulsion
dominates. 
The repulsion mass is not the thermal mass
$M_{\rm thermal}$ (equation \ref{eqn:thermalmass}),
but exceeds it
by a factor of $\sim$$h^{-3/4}$, where $h$ is the
disc aspect ratio.
Growth continues more slowly at $M_{\rm p} > M_{\rm repulsion,visc}$,
with the planet mass increasing beyond $M_{\rm repulsion,visc}$ by up to a factor of $\sim$4 for our parameter choices.

Equation (\ref{eqn:fullsolvisc}) gives an approximate analytic expression
for the planet mass vs.~time during this final repulsion-limited
stage. It predicts that planet masses are of order
$10 \,M_{\rm J}$ by the time the disc dissipates. This result is derived
by assuming the planet accretes
at a rate that scales as $A_{\rm Hill} = 2.2 \, \Omega r^2 \, m^{2/3}$, where $m$ is
the planet-to-star mass ratio;
this prescription is commonly adopted by hydrodynamical
simulations of planet-disc interactions,
and might be appropriate for super-thermal masses. 
If instead of $A_{\rm Hill}$ we
use the empirical formula 
$A_{\rm TW} = 0.29 \, \Omega r^2 m^{4/3}/h^2$ drawn from 2D numerical simulations by \citet{tw_2002}, then 
the mass above which repulsion dominates changes to
$M_{\rm repulsion,visc,TW} \simeq 9 \, M_{\rm J} \,[r / (10 \, {\rm au})]^{3/8}$, nearly twice the value of $M_{\rm repulsion,visc}$
derived using the Hill scaling. Using $A_{\rm TW}$ 
leads to a more extended consumption-dominated growth phase,
and final planet masses larger by order-unity factors
compared to those of the solid curves in Figure \ref{fig:m_final_r_visc}.
Overall, it appears that 
in viscous discs,
planets accrete a not-small fraction
of the disc mass,
which can be many tens of Jupiter masses (\citealt{tab_2017}, their fig.~10; see also \citealt{pmp_2019}). This is in agreement
with \citet{tmt_2020}, who limit giant planet
growth by incorporating photoevaporative mass loss
from the disc.

In inviscid discs, conditions tend to be
repulsion-dominated even at low planet masses. 
Without viscosity or turbulent transport to compete against, 
planetary Lindblad torques carve deep gaps
that are repulsion-dominated even for sub-thermal
planets accreting at the Bondi rate (equation \ref{eqn:sensitive}).
Repulsion-dominated gaps 
are symmetric in the sense that gap contrasts between the outer and inner
discs are the same; accordingly, disc accretion rates are nearly
continuous across the gap (e.g., Figure \ref{fig:m_final_wind_tdisc_b}), which means that most of the disc 
mass is not diverted onto the planet (in the language
of \citealt{tt_2016}, $\dot{M}_{\rm p} = \min (\dot{M}_{\rm hydro}, \dot{M}_{\rm disc}) = \dot{M}_{\rm hydro}$).
Maintaining the disc accretion rate
across a gap demands that the radial accretion velocity within the gap be as large as the gap is deep. Whether such fast inflows are possible,
and whether inviscid gaps can be as deep
as expected from our analytics
(cf.~numerical simulations that
find only shallow gaps;
\citealt{fc_2017,mnp_2019,mnp_2020}),
are unresolved issues.

\begin{figure}  
    \centering
    \includegraphics[width=\linewidth]{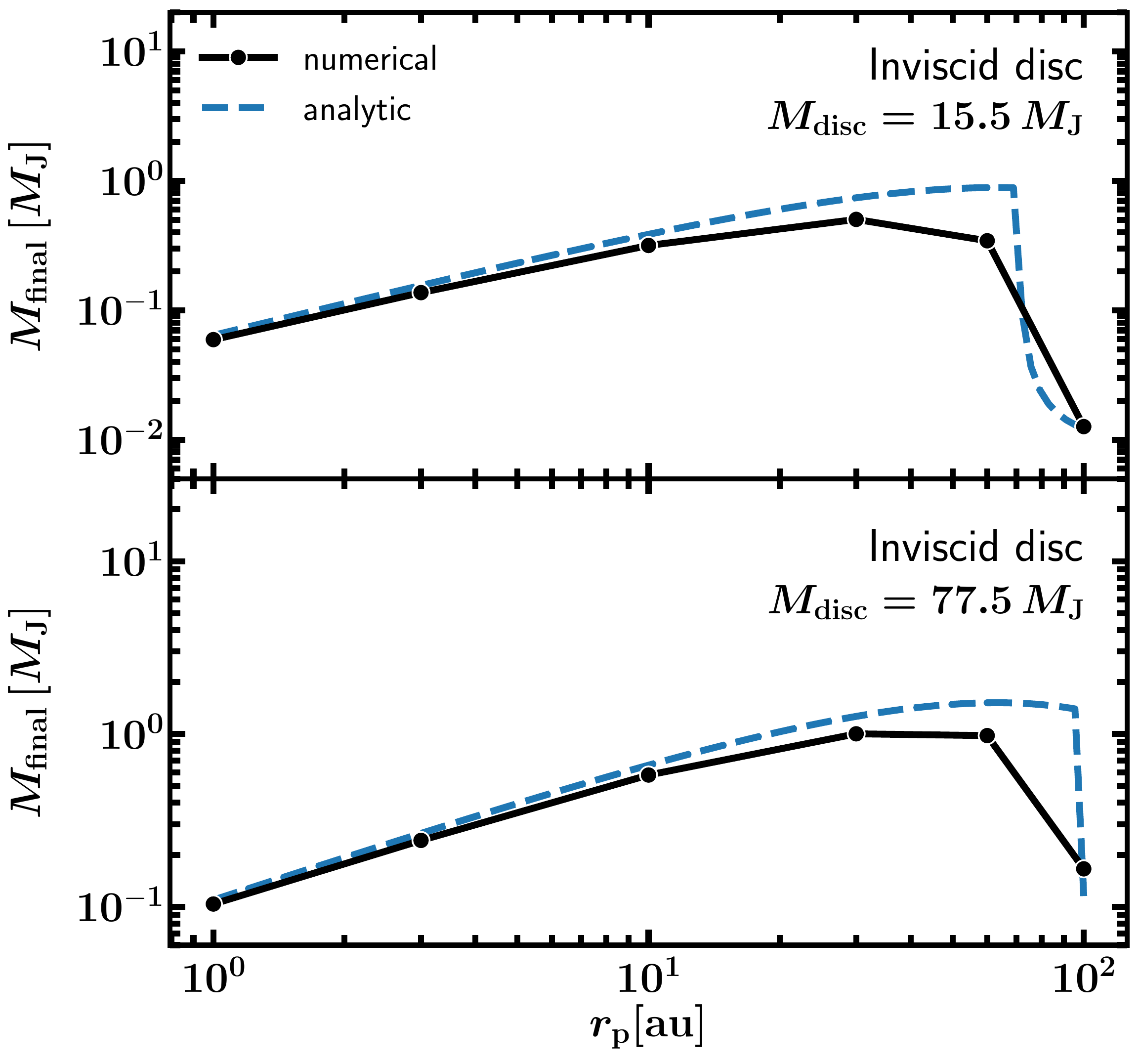}
    \caption{Final planet masses in an inviscid, wind-driven disc
of varying mass (top vs.~bottom panels). Planet masses are
initialized at $0.01 M_{\rm J}$ and grown using the 1D numerical code
of section \ref{sec:numwind}, which uses 
the time-dependent gap contrast of \citet{gc_2019a} to model repulsion,
in a purely advective disc whose height-averaged
radial accretion velocity is $c = -4$ cm/s and exponential
drain-out time is $t_{\rm adv} = 3$ Myr. Points
are plotted after $5 t_{\rm adv} = 15$ Myr. They
mostly respect
equation (\ref{eqn:m_final_wind_full}), which gives
final planet masses grown in repulsion-limited and deep ($\widetilde{B}_{\rm inv}$ > 1) gaps (dashed curve not including the drop-off at the largest
distances).
At $r_{\rm p} \sim 100$ au, the disc has such low density that the planet's initial growth timescale $M_\mathrm{p}/\dot{M}_\mathrm{p}$ is comparable to $t_\mathrm{adv}$; here there are not many doublings before the disc drains away. In this regime the planet does not open a substantial gap ($\widetilde{B}_\mathrm{inv} < 1$) and its final mass can be estimated analytically by integrating $\dot{M}_\mathrm{p} = A_\mathrm{Bondi} \Sigma$ with
$\Sigma$ given by the no-planet solution (\ref{eqn:sol}); the dashed curve is the minimum of the resulting expression (not shown) and \eqref{eqn:m_final_wind_full}.
}
    \label{fig:m_final_r_wind}
\end{figure}

Figure \ref{fig:m_final_r_wind},
analogous to Figure \ref{fig:m_final_r_visc}, shows that final
planet masses in our model inviscid discs 
range between $\sim$0.05 and $1\, M_{\rm J}$, more than
an order-of-magnitude smaller than their viscous disc counterparts. 
For the most part, the masses computed for inviscid
discs using our numerical 1D code are well reproduced
by equation (\ref{eqn:m_final_wind_full}), derived in the
repulsion limit. This formula, which predicts
that final planet masses scale as $M_{\rm disc}^{1/3}$
and $r_{\rm p}^{163/196} \simeq r_{\rm p}^{0.83}$,
is similar to that
derived by Ginzburg \& Chiang (\citeyear{gc_2019a},
their equation 19),\footnote{Our final planet
masses are a factor of $\sim$3 lower than theirs,
a consequence largely of their choice for $h$
which is 50\% larger.} and improves upon it by accounting
for the structure and transport properties of the parent
disc---specifically how the disc may accrete by
shedding angular momentum through a magnetized
surface wind (e.g., \citealt{b_2016}).

Orbital migration in viscous discs
has been shown in numerical simulations 
to enhance $\dot{M}_{\rm p}$ relative to the
migration-free case \citep[e.g.,][]{dk_2017}. 
Including migration would only amplify our
finding that final planet masses in viscous discs
are large, approaching if not well within the regime
of brown dwarfs. Accretion rates should
also increase for planets migrating in inviscid, wind-driven
discs;in 3D, strongly sub-thermal planets
have been shown to migrate inward 
(\citealt{mnp_2020}). We may need 
such enhancements in $\dot{M}_{\rm p}$ to explain,
within an inviscid scenario,
giant planets like our own Jupiter, i.e., 
to bring planet masses up to $1 M_{\rm J}$ at distances
of 1--10 au (Figure \ref{fig:m_final_r_wind}).
On the other hand, sub-Jupiter masses, down to
$\sim$0.1 $M_{\rm J}$ in many cases, are inferred
from ALMA observations of disc gaps (\citealt{zzh_2018}),
and suggest that planets there are strongly repelling
inviscid gas.

The asymmetric gap we computed for the viscous disc
model in Figure \ref{fig:sigma_detailed}
suggests a strong, mostly one-sided migration torque
forcing the planet inward. However, this 
is misleading because our numerical procedure does not spatially resolve the gap,
whose true radial width lies between $H$ (the pressure scale height)
and $r_{\rm p}$ (\citealt{gs_2018}). Most of the migration torque is
exerted by disc gas on the bottoms of gaps, displaced radially from the planet by
$\sim$$\pm H$, 
and here the actual surface density gradients,
and of course the surface density itself, are small (see also
\citealt{kts_2018}).


\subsection{Transitional discs}\label{sec:transitional}
We have shown how a planet accreting
from its parent disc can change the disc's entire
complexion. This make-over is most evident for
a planet that siphons away most of the disc's 
accretion flow---as it can in a viscously diffusing disc---carving out a consumption-limited 
gap that divides a gas-rich outer disc with
surface density $\Sigma_+$
from a gas-poor inner one with surface density
$\Sigma_-$. Transitional
discs have just such an outer/inner structure
(e.g., \citealt{emj_2014}; 
\citealt{dvh_2017}), suggesting that they
represent viscous discs whose inner regions
are cleared by accreting planets
(with dust filtration at the outer gap edge,
and grain growth in the inner disc, 
enhancing the surface density contrast in dust
over gas; \citealt{drz_2012}; \citealt{znd_2012}).

In a viscous disc, a single accreting planet suffices
to deplete the entire disc interior to its orbit.
The 2D single-planet simulations of \citet{znh_2011} 
bear this out; they find an outer vs.~inner
disc contrast of $\Sigma_+/\Sigma_- \sim 10$ 
for a $1 \, M_{\rm J}$ planet that accretes at the Hill
rate from a disc of $h \simeq 0.05$ (their fig.~1, model P1A1). This numerical result agrees with our analytic theory, which predicts
according to equations (\ref{eqn:mdot-0}) and
(\ref{eq:A_B_visc_Hill}) that
\begin{align}
\frac{\Sigma_+}{\Sigma_-} \simeq \frac{A_{\rm Hill}}{3\pi B} \simeq 7 \left( \frac{M_{\rm p}}{M_{\rm J}} \right)^{-4/3} \left( \frac{h}{0.05} \right)^3
\end{align}
for a consumption-dominated and deep gap with $A_{\rm Hill}/(3\pi) > B > \nu$, where $\nu$ is the disc
viscosity.
In steady state, $\Sigma_+/\Sigma_- = \dot{M}_+/\dot{M}_-$,
the ratio of outer-to-inner
disc accretion rates.
A value of $\dot{M}_+/\dot{M}_- \sim 10$,
as we have found
for the above parameters, 
accords with the observation
that the median accretion rate for stars hosting
transitional discs is lower than that of stars hosting
non-transitional discs by a factor of $\sim$10
(\citealt{nsm_2007}; \citealt{kwm_2013}). 
However, the corresponding factor-of-10 reduction
in $\Sigma$ seems too small
to match observed gas depletions in transitional disc cavities.
In the disc studied in CO by \citet{dvh_2017}, the
gas surface density declines by $\sim$$10^3$
from $r = 70$ au to 15 au.
As recognized by Zhu et al.~(\citeyear{znh_2011}; see also \citealt{o_2016}), it is a challenge to
simultaneously explain how disc inner cavities can
be strongly depleted in density while their central stars continue
to accrete at near-normal rates.

This challenge seems more easily met in the repulsion limit,
where deep gaps are carved by planets which alter
the disc accretion flow only modestly---assuming
radial accretion velocities within the gap
are large enough to maintain mass transport rates across it.

The repulsion limit is attained in viscous discs
by planets having $M > M_{\rm repulsion,visc} \simeq 5.3 \, M_{\rm J}\, [r/(10\,{\rm au})]^{9/16}$, or in inviscid discs by
planets having $M > M_{\rm repulsion,inv} \simeq 0.02 \,M_{\rm J}
[r/(10 \,{\rm au})]^{489/392}$.
In both cases, multiple planets with adjoining gaps
would be required to evacuate transition disc cavities
spanning decades in radius---more planets in a viscous scenario
where each gap has a radial width closer to $H$,
and fewer in an inviscid
scenario where each gap is of order $r_{\rm p} > H$ wide
(\citealt{gs_2018}; 
note that widths are not captured by our single-grid-point
treatment of gaps). 
The inviscid picture requires only super-Earth masses and
appeals more, insofar
as observations seem to have already ruled out transitional discs containing
families of super-Jupiters as 
required in the viscous scenario.
Inviscid discs can still accrete, 
either by virtue of magnetized winds
(\citealt{b_2016}; \citealt{wg_2017}),
or by the repulsive torques of their embedded planets
(\citealt{gr_2001}; \citealt{sg_2004};
\citealt{fc_2017}). 

\section*{Acknowledgements} 
We thank Xuening Bai, Jeffrey Fung, Willy Kley, Eve Lee, James Owen, and Hidekazu Tanaka for helpful exchanges. An anonymous referee provided an encouraging report. EC acknowledges NASA grants 80NSSC19K0506 and NNX15AD95G/NEXSS. 
SG is supported by the Heising-Simons Foundation through a 51 Pegasi b Fellowship. MMR and RMC acknowledge support from NSF CAREER grant number AST-1555385.

\section*{Data availability}
The code underlying this article will be shared on reasonable request to the corresponding author.

\bibliographystyle{mnras}
\bibliography{refs} 

\begin{thebibliography}{}
\makeatletter
\relax
\def\mn@urlcharsother{\let\do\@makeother \do\$\do\&\do\#\do\^\do\_\do\%\do\~}
\def\mn@doi{\begingroup\mn@urlcharsother \@ifnextchar [ {\mn@doi@}
  {\mn@doi@[]}}
\def\mn@doi@[#1]#2{\def\@tempa{#1}\ifx\@tempa\@empty \href
  {http://dx.doi.org/#2} {doi:#2}\else \href {http://dx.doi.org/#2} {#1}\fi
  \endgroup}
\def\mn@eprint#1#2{\mn@eprint@#1:#2::\@nil}
\def\mn@eprint@arXiv#1{\href {http://arxiv.org/abs/#1} {{\tt arXiv:#1}}}
\def\mn@eprint@dblp#1{\href {http://dblp.uni-trier.de/rec/bibtex/#1.xml}
  {dblp:#1}}
\def\mn@eprint@#1:#2:#3:#4\@nil{\def\@tempa {#1}\def\@tempb {#2}\def\@tempc
  {#3}\ifx \@tempc \@empty \let \@tempc \@tempb \let \@tempb \@tempa \fi \ifx
  \@tempb \@empty \def\@tempb {arXiv}\fi \@ifundefined
  {mn@eprint@\@tempb}{\@tempb:\@tempc}{\expandafter \expandafter \csname
  mn@eprint@\@tempb\endcsname \expandafter{\@tempc}}}

\bibitem[\protect\citeauthoryear{{Bai}}{{Bai}}{2011}]{b_2011}
{Bai} X.-N.,  2011, \mn@doi [\apj] {10.1088/0004-637X/739/1/50}, \href
  {https://ui.adsabs.harvard.edu/abs/2011ApJ...739...50B} {739, 50}

\bibitem[\protect\citeauthoryear{{Bai}}{{Bai}}{2016}]{b_2016}
{Bai} X.-N.,  2016, \mn@doi [\apj] {10.3847/0004-637X/821/2/80}, \href
  {https://ui.adsabs.harvard.edu/abs/2016ApJ...821...80B} {821, 80}

\bibitem[\protect\citeauthoryear{{Bai} \& {Stone}}{{Bai} \&
  {Stone}}{2013}]{bs_2013}
{Bai} X.-N.,  {Stone} J.~M.,  2013, \mn@doi [\apj]
  {10.1088/0004-637X/769/1/76}, \href
  {https://ui.adsabs.harvard.edu/abs/2013ApJ...769...76B} {769, 76}

\bibitem[\protect\citeauthoryear{{Bai}, {Ye}, {Goodman}  \& {Yuan}}{{Bai}
  et~al.}{2016}]{b_etal_2016}
{Bai} X.-N.,  {Ye} J.,  {Goodman} J.,   {Yuan} F.,  2016, \mn@doi [\apj]
  {10.3847/0004-637X/818/2/152}, \href
  {https://ui.adsabs.harvard.edu/abs/2016ApJ...818..152B} {818, 152}

\bibitem[\protect\citeauthoryear{{Blandford} \& {Payne}}{{Blandford} \&
  {Payne}}{1982}]{bp_1982}
{Blandford} R.~D.,  {Payne} D.~G.,  1982, \mn@doi [\mnras]
  {10.1093/mnras/199.4.883}, \href
  {https://ui.adsabs.harvard.edu/abs/1982MNRAS.199..883B} {199, 883}

\bibitem[\protect\citeauthoryear{{D'Angelo}, {Kley}  \& {Henning}}{{D'Angelo}
  et~al.}{2003}]{dkh_2003}
{D'Angelo} G.,  {Kley} W.,   {Henning} T.,  2003, \mn@doi [\apj]
  {10.1086/367555}, \href
  {https://ui.adsabs.harvard.edu/abs/2003ApJ...586..540D} {586, 540}

\bibitem[\protect\citeauthoryear{{Dong} et~al.,}{{Dong}
  et~al.}{2012}]{drz_2012}
{Dong} R.,  et~al., 2012, \mn@doi [\apj] {10.1088/0004-637X/750/2/161}, \href
  {https://ui.adsabs.harvard.edu/abs/2012ApJ...750..161D} {750, 161}

\bibitem[\protect\citeauthoryear{{Dong} et~al.,}{{Dong}
  et~al.}{2017}]{dvh_2017}
{Dong} R.,  et~al., 2017, \mn@doi [\apj] {10.3847/1538-4357/aa5abf}, \href
  {https://ui.adsabs.harvard.edu/abs/2017ApJ...836..201D} {836, 201}

\bibitem[\protect\citeauthoryear{{Duffell}}{{Duffell}}{2015}]{d_2015}
{Duffell} P.~C.,  2015, \mn@doi [\apjl] {10.1088/2041-8205/807/1/L11}, \href
  {https://ui.adsabs.harvard.edu/abs/2015ApJ...807L..11D} {807, L11}

\bibitem[\protect\citeauthoryear{{Duffell}}{{Duffell}}{2020}]{d_2020}
{Duffell} P.~C.,  2020, \mn@doi [\apj] {10.3847/1538-4357/ab5b0f}, \href
  {https://ui.adsabs.harvard.edu/abs/2020ApJ...889...16D} {889, 16}

\bibitem[\protect\citeauthoryear{{Duffell} \& {Chiang}}{{Duffell} \&
  {Chiang}}{2015}]{dc_2015}
{Duffell} P.~C.,  {Chiang} E.,  2015, \mn@doi [\apj]
  {10.1088/0004-637X/812/2/94}, \href
  {https://ui.adsabs.harvard.edu/abs/2015ApJ...812...94D} {812, 94}

\bibitem[\protect\citeauthoryear{{Duffell} \& {MacFadyen}}{{Duffell} \&
  {MacFadyen}}{2013}]{dm_2013}
{Duffell} P.~C.,  {MacFadyen} A.~I.,  2013, \mn@doi [\apj]
  {10.1088/0004-637X/769/1/41}, \href
  {https://ui.adsabs.harvard.edu/abs/2013ApJ...769...41D} {769, 41}

\bibitem[\protect\citeauthoryear{{Duffell}, {Haiman}, {MacFadyen}, {D'Orazio}
  \& {Farris}}{{Duffell} et~al.}{2014}]{dhm_2014}
{Duffell} P.~C.,  {Haiman} Z.,  {MacFadyen} A.~I.,  {D'Orazio} D.~J.,
  {Farris} B.~D.,  2014, \mn@doi [\apjl] {10.1088/2041-8205/792/1/L10}, \href
  {https://ui.adsabs.harvard.edu/abs/2014ApJ...792L..10D} {792, L10}

\bibitem[\protect\citeauthoryear{{D{\"u}rmann} \& {Kley}}{{D{\"u}rmann} \&
  {Kley}}{2015}]{dk_2015}
{D{\"u}rmann} C.,  {Kley} W.,  2015, \mn@doi [\aap]
  {10.1051/0004-6361/201424837}, \href
  {https://ui.adsabs.harvard.edu/abs/2015A&A...574A..52D} {574, A52}

\bibitem[\protect\citeauthoryear{{D{\"u}rmann} \& {Kley}}{{D{\"u}rmann} \&
  {Kley}}{2017}]{dk_2017}
{D{\"u}rmann} C.,  {Kley} W.,  2017, \mn@doi [\aap]
  {10.1051/0004-6361/201629074}, \href
  {https://ui.adsabs.harvard.edu/abs/2017A&A...598A..80D} {598, A80}

\bibitem[\protect\citeauthoryear{{Espaillat} et~al.,}{{Espaillat}
  et~al.}{2014}]{emj_2014}
{Espaillat} C.,  et~al., 2014, in {Beuther} H.,  {Klessen} R.~S.,  {Dullemond}
  C.~P.,   {Henning} T.,  eds, Protostars and Planets VI. p.~497 (\mn@eprint
  {arXiv} {1402.7103}), \mn@doi{10.2458/azu_uapress_9780816531240-ch022}

\bibitem[\protect\citeauthoryear{{Flaherty} et~al.,}{{Flaherty}
  et~al.}{2017}]{f_etal_2017}
{Flaherty} K.~M.,  et~al., 2017, \mn@doi [\apj] {10.3847/1538-4357/aa79f9},
  \href {https://ui.adsabs.harvard.edu/abs/2017ApJ...843..150F} {843, 150}

\bibitem[\protect\citeauthoryear{{Frank}, {King}  \& {Raine}}{{Frank}
  et~al.}{2002}]{fkr_2002}
{Frank} J.,  {King} A.,   {Raine} D.~J.,  2002, {Accretion Power in
  Astrophysics: Third Edition}

\bibitem[\protect\citeauthoryear{{Fung} \& {Chiang}}{{Fung} \&
  {Chiang}}{2016}]{fc_2016}
{Fung} J.,  {Chiang} E.,  2016, \mn@doi [\apj] {10.3847/0004-637X/832/2/105},
  \href {http://adsabs.harvard.edu/abs/2016ApJ...832..105F} {832, 105}

\bibitem[\protect\citeauthoryear{{Fung} \& {Chiang}}{{Fung} \&
  {Chiang}}{2017}]{fc_2017}
{Fung} J.,  {Chiang} E.,  2017, \mn@doi [\apj] {10.3847/1538-4357/aa6934},
  \href {https://ui.adsabs.harvard.edu/abs/2017ApJ...839..100F} {839, 100}

\bibitem[\protect\citeauthoryear{{Fung}, {Shi}  \& {Chiang}}{{Fung}
  et~al.}{2014}]{fsc_2014}
{Fung} J.,  {Shi} J.-M.,   {Chiang} E.,  2014, \mn@doi [\apj]
  {10.1088/0004-637X/782/2/88}, \href
  {http://adsabs.harvard.edu/abs/2014ApJ...782...88F} {782, 88}

\bibitem[\protect\citeauthoryear{{Fung}, {Zhu}  \& {Chiang}}{{Fung}
  et~al.}{2019}]{fzc_2019}
{Fung} J.,  {Zhu} Z.,   {Chiang} E.,  2019, \mn@doi [\apj]
  {10.3847/1538-4357/ab53da}, \href
  {https://ui.adsabs.harvard.edu/abs/2019ApJ...887..152F} {887, 152}

\bibitem[\protect\citeauthoryear{{Gammie}}{{Gammie}}{1996}]{g_1996}
{Gammie} C.~F.,  1996, \mn@doi [\apj] {10.1086/176735}, \href
  {https://ui.adsabs.harvard.edu/abs/1996ApJ...457..355G} {457, 355}

\bibitem[\protect\citeauthoryear{{Ginzburg} \& {Chiang}}{{Ginzburg} \&
  {Chiang}}{2019a}]{gc_2019a}
{Ginzburg} S.,  {Chiang} E.,  2019a, \mn@doi [\mnras] {10.1093/mnras/stz1322},
  \href {https://ui.adsabs.harvard.edu/abs/2019MNRAS.487..681G} {487, 681}

\bibitem[\protect\citeauthoryear{{Ginzburg} \& {Chiang}}{{Ginzburg} \&
  {Chiang}}{2019b}]{gc_2019b}
{Ginzburg} S.,  {Chiang} E.,  2019b, \mn@doi [\mnras] {10.1093/mnras/stz2901},
  \href {https://ui.adsabs.harvard.edu/abs/2019MNRAS.490.4334G} {490, 4334}

\bibitem[\protect\citeauthoryear{{Ginzburg} \& {Sari}}{{Ginzburg} \&
  {Sari}}{2018}]{gs_2018}
{Ginzburg} S.,  {Sari} R.,  2018, \mn@doi [\mnras] {10.1093/mnras/sty1466},
  \href {https://ui.adsabs.harvard.edu/abs/2018MNRAS.479.1986G} {479, 1986}

\bibitem[\protect\citeauthoryear{{Goldreich} \& {Sari}}{{Goldreich} \&
  {Sari}}{2003}]{gs_2003}
{Goldreich} P.,  {Sari} R.,  2003, \mn@doi [\apj] {10.1086/346202}, \href
  {https://ui.adsabs.harvard.edu/abs/2003ApJ...585.1024G} {585, 1024}

\bibitem[\protect\citeauthoryear{{Goldreich} \& {Tremaine}}{{Goldreich} \&
  {Tremaine}}{1980}]{gt_1980}
{Goldreich} P.,  {Tremaine} S.,  1980, \mn@doi [\apj] {10.1086/158356}, \href
  {https://ui.adsabs.harvard.edu/abs/1980ApJ...241..425G} {241, 425}

\bibitem[\protect\citeauthoryear{{Goldreich} \& {Tremaine}}{{Goldreich} \&
  {Tremaine}}{1982}]{gt_1982}
{Goldreich} P.,  {Tremaine} S.,  1982, \mn@doi [\araa]
  {10.1146/annurev.aa.20.090182.001341}, \href
  {https://ui.adsabs.harvard.edu/abs/1982ARA&A..20..249G} {20, 249}

\bibitem[\protect\citeauthoryear{{Goodman} \& {Rafikov}}{{Goodman} \&
  {Rafikov}}{2001}]{gr_2001}
{Goodman} J.,  {Rafikov} R.~R.,  2001, \mn@doi [\apj] {10.1086/320572}, \href
  {https://ui.adsabs.harvard.edu/abs/2001ApJ...552..793G} {552, 793}

\bibitem[\protect\citeauthoryear{{Hartmann}, {Calvet}, {Gullbring}  \&
  {D'Alessio}}{{Hartmann} et~al.}{1998}]{hcg_1998}
{Hartmann} L.,  {Calvet} N.,  {Gullbring} E.,   {D'Alessio} P.,  1998, \mn@doi
  [\apj] {10.1086/305277}, \href
  {http://adsabs.harvard.edu/abs/1998ApJ...495..385H} {495, 385}

\bibitem[\protect\citeauthoryear{{Kanagawa}, {Muto}, {Tanaka}, {Tanigawa},
  {Takeuchi}, {Tsukagoshi}  \& {Momose}}{{Kanagawa} et~al.}{2015}]{kmt_2015}
{Kanagawa} K.~D.,  {Muto} T.,  {Tanaka} H.,  {Tanigawa} T.,  {Takeuchi} T.,
  {Tsukagoshi} T.,   {Momose} M.,  2015, \mn@doi [\apjl]
  {10.1088/2041-8205/806/1/L15}, \href
  {https://ui.adsabs.harvard.edu/abs/2015ApJ...806L..15K} {806, L15}

\bibitem[\protect\citeauthoryear{{Kanagawa}, {Tanaka}  \&
  {Szuszkiewicz}}{{Kanagawa} et~al.}{2018}]{kts_2018}
{Kanagawa} K.~D.,  {Tanaka} H.,   {Szuszkiewicz} E.,  2018, \mn@doi [\apj]
  {10.3847/1538-4357/aac8d9}, \href
  {https://ui.adsabs.harvard.edu/abs/2018ApJ...861..140K} {861, 140}

\bibitem[\protect\citeauthoryear{{Kim} et~al.,}{{Kim} et~al.}{2013}]{kwm_2013}
{Kim} K.~H.,  et~al., 2013, \mn@doi [\apj] {10.1088/0004-637X/769/2/149}, \href
  {https://ui.adsabs.harvard.edu/abs/2013ApJ...769..149K} {769, 149}

\bibitem[\protect\citeauthoryear{{Kimmig}, {Dullemond}  \& {Kley}}{{Kimmig}
  et~al.}{2020}]{kdk_2020}
{Kimmig} C.~N.,  {Dullemond} C.~P.,   {Kley} W.,  2020, \mn@doi [\aap]
  {10.1051/0004-6361/201936412}, \href
  {https://ui.adsabs.harvard.edu/abs/2020A&A...633A...4K} {633, A4}

\bibitem[\protect\citeauthoryear{{Kley} \& {Dirksen}}{{Kley} \&
  {Dirksen}}{2006}]{kd_2006}
{Kley} W.,  {Dirksen} G.,  2006, \mn@doi [\aap] {10.1051/0004-6361:20053914},
  \href {https://ui.adsabs.harvard.edu/abs/2006A&A...447..369K} {447, 369}

\bibitem[\protect\citeauthoryear{{Kley} \& {Nelson}}{{Kley} \&
  {Nelson}}{2012}]{kn_2012}
{Kley} W.,  {Nelson} R.~P.,  2012, \mn@doi [\araa]
  {10.1146/annurev-astro-081811-125523}, \href
  {https://ui.adsabs.harvard.edu/abs/2012ARA&A..50..211K} {50, 211}

\bibitem[\protect\citeauthoryear{{Lee}}{{Lee}}{2019}]{lee_2019}
{Lee} E.~J.,  2019, \mn@doi [\apj] {10.3847/1538-4357/ab1b40}, \href
  {https://ui.adsabs.harvard.edu/abs/2019ApJ...878...36L} {878, 36}

\bibitem[\protect\citeauthoryear{{Lin} \& {Papaloizou}}{{Lin} \&
  {Papaloizou}}{1986}]{lp_1986}
{Lin} D.~N.~C.,  {Papaloizou} J.,  1986, \mn@doi [\apj] {10.1086/164653}, \href
  {https://ui.adsabs.harvard.edu/abs/1986ApJ...309..846L} {309, 846}

\bibitem[\protect\citeauthoryear{{Lubow} \& {D'Angelo}}{{Lubow} \&
  {D'Angelo}}{2006}]{ld_2006}
{Lubow} S.~H.,  {D'Angelo} G.,  2006, \mn@doi [\apj] {10.1086/500356}, \href
  {https://ui.adsabs.harvard.edu/abs/2006ApJ...641..526L} {641, 526}

\bibitem[\protect\citeauthoryear{{Lynden-Bell} \& {Pringle}}{{Lynden-Bell} \&
  {Pringle}}{1974}]{lp_1974}
{Lynden-Bell} D.,  {Pringle} J.~E.,  1974, \mn@doi [\mnras]
  {10.1093/mnras/168.3.603}, \href
  {https://ui.adsabs.harvard.edu/abs/1974MNRAS.168..603L} {168, 603}

\bibitem[\protect\citeauthoryear{{McNally}, {Nelson}, {Paardekooper}  \&
  {Ben{\'\i}tez-Llambay}}{{McNally} et~al.}{2019}]{mnp_2019}
{McNally} C.~P.,  {Nelson} R.~P.,  {Paardekooper} S.-J.,
  {Ben{\'\i}tez-Llambay} P.,  2019, \mn@doi [\mnras] {10.1093/mnras/stz023},
  \href {https://ui.adsabs.harvard.edu/abs/2019MNRAS.484..728M} {484, 728}

\bibitem[\protect\citeauthoryear{{McNally}, {Nelson}, {Paardekooper},
  {Ben{\'\i}tez-Llambay}  \& {Gressel}}{{McNally} et~al.}{2020}]{mnp_2020}
{McNally} C.~P.,  {Nelson} R.~P.,  {Paardekooper} S.-J.,
  {Ben{\'\i}tez-Llambay} P.,   {Gressel} O.,  2020, \mn@doi [\mnras]
  {10.1093/mnras/staa576}, \href
  {https://ui.adsabs.harvard.edu/abs/2020MNRAS.493.4382M} {493, 4382}

\bibitem[\protect\citeauthoryear{{Morbidelli}, {Szul{\'a}gyi}, {Crida}, {Lega},
  {Bitsch}, {Tanigawa}  \& {Kanagawa}}{{Morbidelli} et~al.}{2014}]{msc_2014}
{Morbidelli} A.,  {Szul{\'a}gyi} J.,  {Crida} A.,  {Lega} E.,  {Bitsch} B.,
  {Tanigawa} T.,   {Kanagawa} K.,  2014, \mn@doi [\icarus]
  {10.1016/j.icarus.2014.01.010}, \href
  {http://adsabs.harvard.edu/abs/2014Icar..232..266M} {232, 266}

\bibitem[\protect\citeauthoryear{{Muley}, {Fung}  \& {van der Marel}}{{Muley}
  et~al.}{2019}]{mfv_2019}
{Muley} D.,  {Fung} J.,   {van der Marel} N.,  2019, \mn@doi [\apjl]
  {10.3847/2041-8213/ab24d0}, \href
  {https://ui.adsabs.harvard.edu/abs/2019ApJ...879L...2M} {879, L2}

\bibitem[\protect\citeauthoryear{{Najita}, {Strom}  \& {Muzerolle}}{{Najita}
  et~al.}{2007}]{nsm_2007}
{Najita} J.~R.,  {Strom} S.~E.,   {Muzerolle} J.,  2007, \mn@doi [\mnras]
  {10.1111/j.1365-2966.2007.11793.x}, \href
  {https://ui.adsabs.harvard.edu/abs/2007MNRAS.378..369N} {378, 369}

\bibitem[\protect\citeauthoryear{{Owen}}{{Owen}}{2016}]{o_2016}
{Owen} J.~E.,  2016, \mn@doi [\pasa] {10.1017/pasa.2016.2}, \href
  {https://ui.adsabs.harvard.edu/abs/2016PASA...33....5O} {33, e005}

\bibitem[\protect\citeauthoryear{{Papaloizou}, {Nelson}  \&
  {Masset}}{{Papaloizou} et~al.}{2001}]{pnm_2001}
{Papaloizou} J.~C.~B.,  {Nelson} R.~P.,   {Masset} F.,  2001, \mn@doi [\aap]
  {10.1051/0004-6361:20000011}, \href
  {https://ui.adsabs.harvard.edu/abs/2001A&A...366..263P} {366, 263}

\bibitem[\protect\citeauthoryear{{Perez-Becker} \& {Chiang}}{{Perez-Becker} \&
  {Chiang}}{2011}]{pc_2011}
{Perez-Becker} D.,  {Chiang} E.,  2011, \mn@doi [\apj]
  {10.1088/0004-637X/727/1/2}, \href
  {https://ui.adsabs.harvard.edu/abs/2011ApJ...727....2P} {727, 2}

\bibitem[\protect\citeauthoryear{{Pinte}, {Dent}, {M{\'e}nard}, {Hales},
  {Hill}, {Cortes}  \& {de Gregorio-Monsalvo}}{{Pinte}
  et~al.}{2016}]{p_etal_2016}
{Pinte} C.,  {Dent} W.~R.~F.,  {M{\'e}nard} F.,  {Hales} A.,  {Hill} T.,
  {Cortes} P.,   {de Gregorio-Monsalvo} I.,  2016, \mn@doi [\apj]
  {10.3847/0004-637X/816/1/25}, \href
  {https://ui.adsabs.harvard.edu/abs/2016ApJ...816...25P} {816, 25}

\bibitem[\protect\citeauthoryear{{Powell}, {Murray-Clay}, {P{\'e}rez},
  {Schlichting}  \& {Rosenthal}}{{Powell} et~al.}{2019}]{pmp_2019}
{Powell} D.,  {Murray-Clay} R.,  {P{\'e}rez} L.~M.,  {Schlichting} H.~E.,
  {Rosenthal} M.,  2019, \mn@doi [\apj] {10.3847/1538-4357/ab20ce}, \href
  {https://ui.adsabs.harvard.edu/abs/2019ApJ...878..116P} {878, 116}

\bibitem[\protect\citeauthoryear{Press, Teukolsky, Vetterling  \&
  Flannery}{Press et~al.}{2007}]{ptv_2007}
Press W.~H.,  Teukolsky S.~A.,  Vetterling W.~T.,   Flannery B.~P.,  2007,
  Numerical Recipes 3rd Edition: The Art of Scientific Computing, 3 edn.
Cambridge University Press, New York, NY, USA

\bibitem[\protect\citeauthoryear{{Sari} \& {Goldreich}}{{Sari} \&
  {Goldreich}}{2004}]{sg_2004}
{Sari} R.,  {Goldreich} P.,  2004, \mn@doi [\apjl] {10.1086/421080}, \href
  {https://ui.adsabs.harvard.edu/abs/2004ApJ...606L..77S} {606, L77}

\bibitem[\protect\citeauthoryear{{Shakura} \& {Sunyaev}}{{Shakura} \&
  {Sunyaev}}{1973}]{ss_alpha}
{Shakura} N.~I.,  {Sunyaev} R.~A.,  1973, \aap, \href
  {http://adsabs.harvard.edu/abs/1973A%26A....24..337S} {24, 337}

\bibitem[\protect\citeauthoryear{{Szul{\'a}gyi}, {Morbidelli}, {Crida}  \&
  {Masset}}{{Szul{\'a}gyi} et~al.}{2014}]{smc_2014}
{Szul{\'a}gyi} J.,  {Morbidelli} A.,  {Crida} A.,   {Masset} F.,  2014, \mn@doi
  [\apj] {10.1088/0004-637X/782/2/65}, \href
  {http://adsabs.harvard.edu/abs/2014ApJ...782...65S} {782, 65}

\bibitem[\protect\citeauthoryear{{Tanaka}, {Murase}  \& {Tanigawa}}{{Tanaka}
  et~al.}{2020}]{tmt_2020}
{Tanaka} H.,  {Murase} K.,   {Tanigawa} T.,  2020, \mn@doi [\apj]
  {10.3847/1538-4357/ab77af}, \href
  {https://ui.adsabs.harvard.edu/abs/2020ApJ...891..143T} {891, 143}

\bibitem[\protect\citeauthoryear{{Tanigawa} \& {Tanaka}}{{Tanigawa} \&
  {Tanaka}}{2016}]{tt_2016}
{Tanigawa} T.,  {Tanaka} H.,  2016, \mn@doi [\apj]
  {10.3847/0004-637X/823/1/48}, \href
  {https://ui.adsabs.harvard.edu/abs/2016ApJ...823...48T} {823, 48}

\bibitem[\protect\citeauthoryear{{Tanigawa} \& {Watanabe}}{{Tanigawa} \&
  {Watanabe}}{2002}]{tw_2002}
{Tanigawa} T.,  {Watanabe} S.-i.,  2002, \mn@doi [\apj] {10.1086/343069}, \href
  {https://ui.adsabs.harvard.edu/abs/2002ApJ...580..506T} {580, 506}

\bibitem[\protect\citeauthoryear{{Tripathi}, {Andrews}, {Birnstiel}  \&
  {Wilner}}{{Tripathi} et~al.}{2017}]{tab_2017}
{Tripathi} A.,  {Andrews} S.~M.,  {Birnstiel} T.,   {Wilner} D.~J.,  2017,
  \mn@doi [\apj] {10.3847/1538-4357/aa7c62}, \href
  {https://ui.adsabs.harvard.edu/abs/2017ApJ...845...44T} {845, 44}

\bibitem[\protect\citeauthoryear{{Wang} \& {Goodman}}{{Wang} \&
  {Goodman}}{2017}]{wg_2017}
{Wang} L.,  {Goodman} J.~J.,  2017, \mn@doi [\apj]
  {10.3847/1538-4357/835/1/59}, \href
  {https://ui.adsabs.harvard.edu/abs/2017ApJ...835...59W} {835, 59}

\bibitem[\protect\citeauthoryear{{Wang} et~al.,}{{Wang}
  et~al.}{2018}]{wgd_2018}
{Wang} J.~J.,  et~al., 2018, \mn@doi [\aj] {10.3847/1538-3881/aae150}, \href
  {https://ui.adsabs.harvard.edu/abs/2018AJ....156..192W} {156, 192}

\bibitem[\protect\citeauthoryear{{Zhang} et~al.,}{{Zhang}
  et~al.}{2018}]{zzh_2018}
{Zhang} S.,  et~al., 2018, \mn@doi [\apjl] {10.3847/2041-8213/aaf744}, \href
  {https://ui.adsabs.harvard.edu/abs/2018ApJ...869L..47Z} {869, L47}

\bibitem[\protect\citeauthoryear{{Zhu}, {Nelson}, {Hartmann}, {Espaillat}  \&
  {Calvet}}{{Zhu} et~al.}{2011}]{znh_2011}
{Zhu} Z.,  {Nelson} R.~P.,  {Hartmann} L.,  {Espaillat} C.,   {Calvet} N.,
  2011, \mn@doi [\apj] {10.1088/0004-637X/729/1/47}, \href
  {https://ui.adsabs.harvard.edu/abs/2011ApJ...729...47Z} {729, 47}

\bibitem[\protect\citeauthoryear{{Zhu}, {Nelson}, {Dong}, {Espaillat}  \&
  {Hartmann}}{{Zhu} et~al.}{2012}]{znd_2012}
{Zhu} Z.,  {Nelson} R.~P.,  {Dong} R.,  {Espaillat} C.,   {Hartmann} L.,  2012,
  \mn@doi [\apj] {10.1088/0004-637X/755/1/6}, \href
  {https://ui.adsabs.harvard.edu/abs/2012ApJ...755....6Z} {755, 6}

\makeatother
\end{thebibliography}



\appendix

\section{Analytic Steady-State Solution for $\Sigma$ and $\dot{M}$ For Viscous Disc with Planet }\label{sec:analy_ss}

In this appendix we provide an analytic expression for the surface density profile of a disc with
an embedded planet. Our derivation here is more careful than our order-of-magnitude sketch in section \ref{sec:oom}, and similar to that presented in Lubow \& D'Angelo (\citeyear{ld_2006}, their section 2.4), with a couple of differences:  we reduce the surface density at the planet's location by a factor $1 + B/\nu$ to account for repulsive Lindblad torques (see section \ref{sec:oom}), and we express our solution in terms of the surface density at infinity as opposed to the surface density at the planet's location. 

Using the same notation as in section \ref{sec:oom}, and neglecting for the moment
the Lindblad torque, the equations of mass and angular momentum conservation with a mass sink at $r = r_\mathrm{p}$ read
\begin{align}
\frac{1}{r}\frac{d\left( \mu u_r r / \nu\right)}{dr} & =-\frac{\dot{M}_\mathrm{p}}{2\pi r}\delta\left(r-r_\mathrm{p}\right) \label{eq:mass_cons_ss}\\
\frac{r^{2}\Omega \mu u_r}{\nu} & \label{eq:ang_cons_ss} =-\frac{d}{dr}\left(3\mu\Omega r^{2}\right)
\end{align}
where $u_r$ is the radial velocity and $\mu \equiv \Sigma \nu$.   
Equation \eqref{eq:mass_cons_ss} indicates that the mass flow rate $\dot{M}_+ = -2\pi \mu_+ u_r r/\nu$ is spatially constant in regions exterior to the planet's orbit (the outer disc), and likewise for $\dot{M}_-$ in regions interior to the planet's orbit (the inner disc): 
\begin{align} \label{eq:m_dot_cons}
    \dot{M}_- = \dot{M}_+ - A \frac{\mu_\mathrm{p}}{\nu_\mathrm{p}}
\end{align} 
where we have used $\dot{M}_{\rm p} = A \mu_{\rm p}/\nu_{\rm p}$ and $\nu_\mathrm{p} \equiv \nu \left( r_\mathrm{p} \right)$. Since $\dot{M}_-$ and $\dot{M}_+$ are constants, equation \eqref{eq:ang_cons_ss} can be solved to yield 
\begin{align}
    3 \pi \mu_\pm(r) = \dot{M_\pm} + \frac{C_\pm}{\sqrt{r}}
\end{align}
where $C_\pm$ are integration constants. For the inner disc we use the boundary condition $\mu_-(r_\star) = 0$, whence
\begin{align} \label{eq:mu_sol}
3\pi\mu_{-}\left(r\right)=\dot{M}_{-}\left(1-\sqrt{\frac{r_\star}{r}}\right) \,.
\end{align}
Following our treatment in the main text,
we encode the planetary gap caused by Lindblad
torques at a sub-grid level, i.e., we force
the surface density at the planet's location
to be depleted relative to the surface density just interior
to the planet according to
\begin{align} \label{eq:inner_mu_p}
\mu_{\rm p} = \mu_-(r_{\rm p}) \left( 1 + B/\nu_{\rm p} \right)^{-1}
\end{align}
where subscript $p$ denotes the planet's location.
For the outer disc, we fix the surface density at infinity, $\mu(\infty) = \mu_\infty$, so that
\begin{equation}
\dot{M}_+ = 3 \pi \mu_\infty \,.
\end{equation}
Then from
equations (\ref{eq:m_dot_cons}), (\ref{eq:mu_sol}), and (\ref{eq:inner_mu_p}) 
we have
\begin{align} \label{eq:ss_mup}
\frac{\mu_\mathrm{p}}{\mu_\infty}=\frac{1}{\frac{1}{1-\sqrt{r_\star/r_\mathrm{p}}} + \left( \frac{A}{3 \pi} + \frac{B}{1-\sqrt{r_\star/r_\mathrm{p}}}\right)/\nu_\mathrm{p}}
\end{align}
which can be compared to equation (\ref{eqn:oom2}).
We may also solve for
\begin{align}
    \frac{\dot{M}_\mathrm{p}}{\dot{M}_+} &= \frac{A/(3\pi \nu_\mathrm{p})}{\frac{1}{1-\sqrt{r_\star/r_\mathrm{p}}} + \left(\frac{A}{3 \pi} + \frac{B}{1-\sqrt{r_\star/r_\mathrm{p}}}\right)/\nu_\mathrm{p}}\\
    \frac{\dot{M}_-}{\dot{M}_+} &= \frac{1 + B/\nu_\mathrm{p}}{1 + \left[ \frac{A}{3 \pi} \left(1-\sqrt{\frac{r_\star}{r_\mathrm{p}}}\right) + B \right]/\nu_\mathrm{p}}
\end{align}
which can be compared to equations (\ref{eq:m_dot_ratio}) and (\ref{eqn:mdot-0}).
Finally, stitching the outer disc solution to
the inner disc solution implies
$\mu_+(r_{\rm p}) = \mu_-(r_{\rm p}) = \mu_{\rm p} (1+B/\nu_{\rm p})$ and
\begin{align} \label{eq:plus}
    3 \pi \mu_+(r) = 3\pi \mu_\infty \left[1 - \sqrt{\frac{r_\mathrm{p}}{r}} \left(1-\frac{\mu_\mathrm{p}\left( 1 + B/\nu_\mathrm{p} \right)}{\mu_\infty} \right) \right] \,.
\end{align}
The equations above 
mirror the results in section \ref{sec:oom}, with the 
addition of a factor of $3 \pi$ (see section \ref{sec:num}) and 
the factor of $1-\sqrt{r_\star/r_\mathrm{p}}$ which
accounts for the star's ability to divert material from the planet.

\begin{figure}  
    \centering 
    \includegraphics[width=\linewidth]{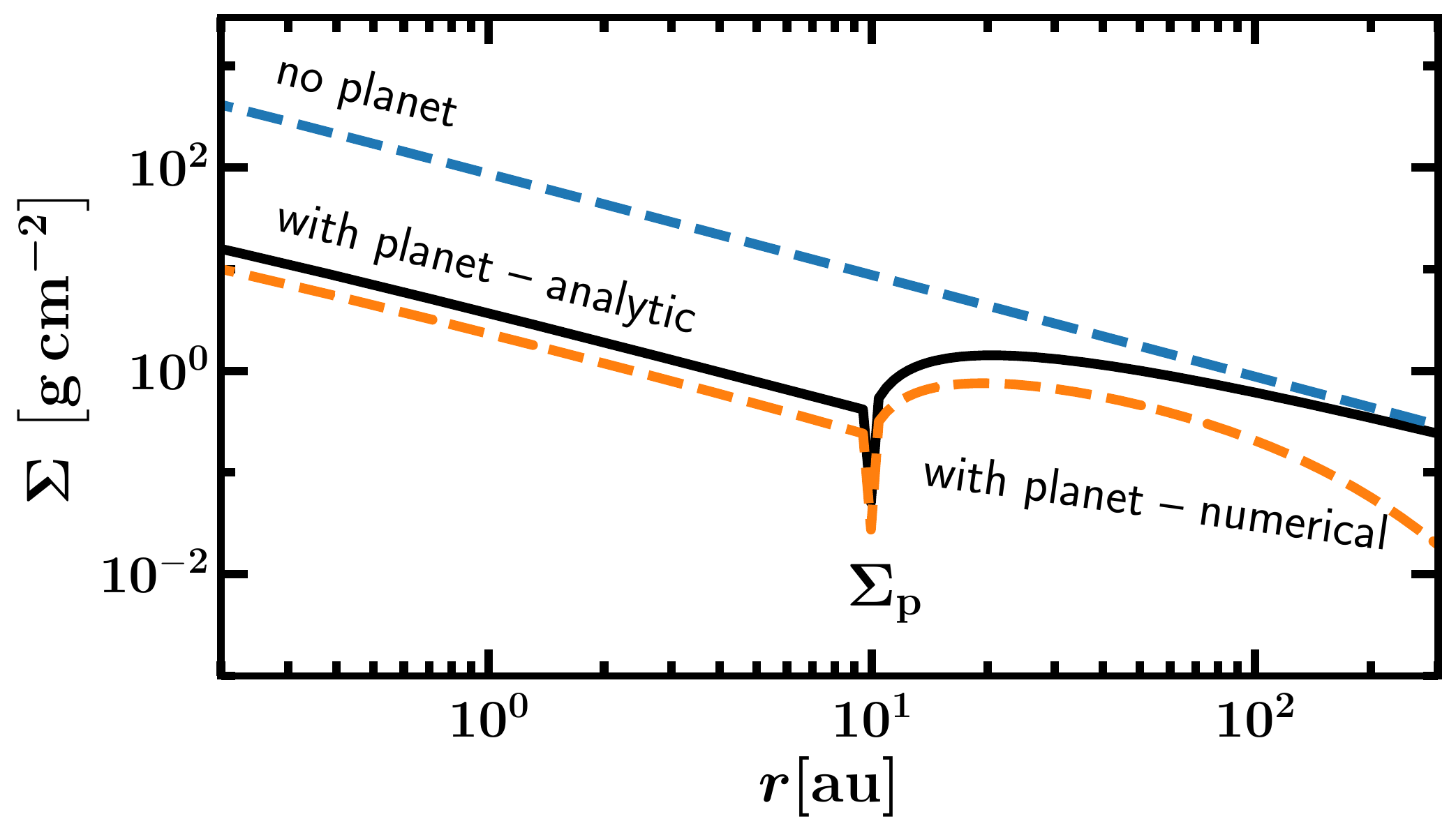}
    \caption{Analytic solution (black solid curve) for the surface density
    profile of a viscous disc perturbed by a planet,
    as given by equations (\ref{eq:mu_sol}), (\ref{eq:inner_mu_p}) and (\ref{eq:plus}),
    using parameters as close as possible to those
    used in the top panel of Figure \ref{fig:sigma_detailed},
    whose numerical result is overlaid here for comparison (orange dashed curve).
    For our analytic unperturbed ``no planet'' disc (blue dashed curve) we use
a power law of slope -1 and normalization at 1 au equal
to the corresponding ``no planet'' curve in Figure \ref{fig:sigma_detailed}. 
The $A$ and $B$ coefficients are taken from equations 
\eqref{eqn:bondi} and \eqref{eqn:kanagawa} for 
$M_\mathrm{p} = 0.3 \,M_\mathrm{J}$. The differences between the analytic and numeric curves mainly arise from the behaviour of the outermost
disc near the turn-around ``transition radius'' (\citealt{lp_1974}; \citealt{hcg_1998}). This transition radius, which varies with time, does not appear in our steady-state solution.}
    \label{fig:sig_analy_sol}
\end{figure}

In Figure \ref{fig:sig_analy_sol} we plot equations (\ref{eq:mu_sol}), (\ref{eq:inner_mu_p}) and  (\ref{eq:plus}),
adopting parameters as close as possible to those
used in the top panel of Figure \ref{fig:sigma_detailed} so that
we may compare the numerical result there to
the analytic result here (see caption to
Figure \ref{fig:sig_analy_sol} for details).

\section{Magnetized winds and disc accretion}\label{sec:winds}

We motivate here our simple, constant accretion velocity
model for a wind-driven disc using the numerical
simulations of Bai and collaborators.
From continuity (equations 1, 6, and 9 of \citealt{b_2016}),
\begin{align} \label{continuity}
\frac{\partial \Sigma}{\partial t} &= +\frac{1}{2\pi r} \frac{\partial 
  \dot{M}_{\rm disc}} {\partial r} - \frac{1}{2\pi r} \frac{\partial 
  \dot{M}_{\rm wind}}{\partial r} \nonumber \\
&= + \frac{1}{2\pi r}
\frac{\partial}{\partial r} \left[ 2(\lambda-1) r \frac{\partial \dot{M}_{\rm wind}}{\partial r} \right] 
 - \frac{1}{2\pi r} \frac{\partial 
  \dot{M}_{\rm wind}}{\partial r} 
\end{align}
where 
\begin{align}
\dot{M}_{\rm wind} (r) = \int^r \frac{\partial \dot{M}_{\rm
  wind}}{\partial r} dr
\end{align}
is the cumulative rate at which mass is carried
to infinity by the wind (integrated over the disc within $r$). From equation (20) of \citet{b_etal_2016},
\begin{equation}\label{eqn:rhov}
\frac{\partial \dot{M}_{\rm wind}}{\partial  r} = 2\pi r \rho_0 u_{\rm p0} 
\end{equation}
where $\rho_0$ and $u_{\rm p0}$ are the volumetric mass density
and poloidal velocity of the wind where it is launched,
near the disc surface. 
All quantities subscripted with 0 are evaluated
at the wind base $(r_0,z_0)$.

The disc accretion rate
\begin{equation} \label{eqn:usual}
\dot{M}_{\rm disc} \equiv -2\pi \Sigma r u_r = 2(\lambda-1) r \frac{\partial \dot{M}_{\rm wind}}{\partial r}
\end{equation}
for surface density
$\Sigma$ and radial velocity $u_r$ is identical in  definition to the variable $\dot{M}_{\rm disc}$
used throughout our paper. Unlike $\dot{M}_{\rm wind}$,
$\dot{M}_{\rm disc}$ is not a cumulative quantity,
but measures the mass crossing a circle of radius $r$ per unit time,
and uses a sign convention such that $\dot{M}_{\rm disc} > 0$ for $u_r <0$.


Disc accretion by a wind hinges on the ``magnetic lever arm''
\begin{equation}
\lambda = (r_{\rm A}/r_0)^2 \label{eqn:lambda}
\end{equation}
where $r_{\rm A}$ is the Alfv\'en radius for the wind streamline running through $r_0$.
A lever arm $\lambda > 1$
enables $\dot{M}_{\rm disc} > 0$ by having the
wind carry away more specific angular momentum
than the Keplerian disc has at $r_0$. 
The fiducial wind
model of Bai (\citeyear{b_2016}, their fig.~2) has $(\lambda-1)$ ranging from
$\sim$30 at $r = 0.3$ AU to $\sim$2
at 30 AU; therefore the first term in (\ref{continuity})
dominates the second term by a factor of
order $2(\lambda-1) \sim 4$--60. Only the first term is modeled in our paper.

\citet{b_2016} and the magnetized disc wind literature dating back to \citet{bp_1982} parameterize the wind mass-loss rate in terms of the
dimensionless mass loading parameter
\begin{equation}
\mu = \frac{\omega r_0}{B_{\rm p0}} \times k = \frac{\omega
  r_0}{B_{\rm p0}} \times \frac{4\pi \rho u_{\rm p}}{B_{\rm p}} 
\end{equation}
where $k$ is the ratio of poloidal mass flux to poloidal
field strength $B_{\rm p}$
($k$ is constant along a magnetic field line), and $\omega$  
is the angular velocity of a field line, approximately
equal to the Keplerian frequency $\Omega_{\rm K}$ at $r_0$. 
Note that $\mu$ (not to be confused with $\mu$
in Appendix \ref{sec:analy_ss}) 
varies with $r$ from field line to field line. Evaluating $\mu$
at the wind base, we rewrite (\ref{eqn:rhov}) as
\begin{equation}
\frac{\partial \dot{M}_{\rm wind}}{\partial r} = \frac{\mu B_{\rm p0}^2}{2\omega} 
\end{equation}
(\citealt{b_etal_2016}, equation 21). Now parameterize
$B_{\rm p0}$ in terms of the midplane plasma beta:
\begin{equation}
\beta_0 = \frac{8\pi}{B_{\rm p0}^2} \frac{\Sigma k_{\rm B} T}{\sqrt{2\pi} \overline{m} H }
\end{equation}
where $k_{\rm B}$ is Boltzmann's constant, $T$ is the disc
temperature, $H = c_{\rm s}/\Omega_{\rm K}$ is the disc scale height,
$c_{\rm s} = \sqrt{k_{\rm B}T/\overline{m}}$ is the gas sound speed, and $\overline{m}$ 
is the mean molecular weight. Then
\begin{equation} \label{eight}
\frac{\partial \dot{M}_{\rm wind}}{\partial  r} =  \frac{\sqrt{8\pi}k_{\rm B}}{\beta_0 \overline{m}} \frac{T}{H} \frac{\mu \Sigma}{\omega} = \frac{\sqrt{8\pi k_{\rm B}}}{\sqrt{\overline{m}}\beta_0} T^{1/2} \mu  \Sigma \,.
\end{equation}
Combine (\ref{eight}) and (\ref{eqn:usual}) to find
\begin{equation} \label{eqn:end}
u_r = -\sqrt{\frac{8}{\pi}} \frac{\sqrt{k_{\rm B}}}{\sqrt{\overline{m}}\beta_0} T^{1/2} \mu (\lambda-1) \sim -\frac{\mu(\lambda-1)}{\beta_0} c_{\rm s} \,.
\end{equation}
In the fiducial model of Bai (\citeyear{b_2016}, see their fig.~2),
$\mu$ increases from $\sim$0.06 at $r = 0.3$ AU
to $\sim$4 at 30 AU, and $(\lambda-1)$ decreases
from $\sim$30 to $\sim$2 over the same range; therefore the product
$\mu(\lambda-1)$ increases from $\sim$2 to $\sim$8,
scaling roughly as $r^{0.3}$.
Their model temperature 
scales as $T \propto r^{-1/2}$;
therefore the combination $T^{1/2}\mu (\lambda-1)$
is nearly constant with $r$. Assuming it to be constant implies from (\ref{eqn:end}) that $u_r$ is similarly constant (cf.~\citealt{kdk_2020}),
if $\beta_0$ is constant:
\begin{equation}
u_r \sim -4 \left( \frac{10^5}{\beta_0} \right) {\rm cm}/{\rm s} \,.
\end{equation}
Taking $\beta_0$ to be a strict
constant corresponds to a model intermediate between
the conserved-flux model of \citet{b_2016} (dashed line in the right panel
of their fig.~5)
and their flux-proportional-to-mass model (solid line).
Using their initial fiducial $\beta_0 = 10^5$ 
implies the disc at $r = 30$ AU drains out in
$r/|u_r| \simeq 3$ Myr. 


\bsp	
\label{lastpage}
\end{document}